\documentclass[pdflatex,sn-mathphys-num]{sn-jnl}

\usepackage{graphicx,adjustbox,rotating}%
\usepackage{multirow}%
\usepackage{amsmath,amssymb,amsfonts}%
\newcommand{\nbooks}{42}
\newcommand{\ngbooks}{17}
\newcommand{\nugbooks}{25}
\newcommand{\nwikibooks}{18}
\newcommand{\nmodernbooks}{27}
\newcommand{\nmillbooks}{17}
\newcommand{\noldbooks}{15}
\newcommand{\ket}[1]{\left| #1 \right>} 

\begin{document}

\title[How to be an orthodox quantum mechanic]{How to be an orthodox quantum mechanic}


\author{\fnm{Geoff} \sur{Beck}}\email{geoffrey.beck@wits.ac.za}

\affil{\orgdiv{School of Physics}, \orgname{University of Witwatersrand}, \orgaddress{\street{1 Jan Smuts Avenue}, \city{Johannesburg}, \postcode{WITS-2050}, \state{Gauteng}, \country{South Africa}}}


\abstract{This work sets out to answer a single question: what is the orthodox interpretation of quantum mechanics? However, we adopt a different approach to that normally used. Rather than surveying physicists, or poring over the precise details of the thoughts of Bohr and Heisenberg, we review a collection of {\nbooks} textbooks on quantum mechanics, encompassing the most popular and prominent works of this nature. We then gauge their response to 13 propositions to build up a picture of exactly what is believed by an orthodox quantum mechanic. We demonstrate that this orthodoxy has many aspects of Copenhagen-like viewpoints, but also shows some interesting emerging deviations. Moreover, it is more nuanced than some reductive characterisations that condense the orthodoxy down to the ontological primacy of the quantum state. The revealed orthodoxy has three consistent pillars: measurement inherently disturbs quantum states, these states refer to individual instances, not ensembles, and quantum systems do not have definite properties prior to measurement. More fully, it entails that individual particles exist in wave-like super-positions and present particle behaviours only when forced to by outside influences. The act of measuring such a system inherently induces random changes in its state, manifesting as a form of measurement error that corresponds to the uncertainty principle. This implies that measurement does not reveal underlying values of quantum properties.}

\maketitle

\section{Introduction}\label{sec1}

How do scientists understand quantum mechanics? The most popular interpretation of quantum mechanics (QM) is usually found in surveys to be ``Copenhagen interpretation''~\cite{sivasundaram2016surveyingattitudesphysicistsconcerning,survey2025}. However, the definition of this viewpoint is notoriously vague~\cite{hanson1959,stapp1972,Jammer1974-JAMTPO-10,Home1997}, its characterisation tending to descend into hyperfine readings of Bohr and Heisenberg~\cite{stapp1972}.

Max Jammer sets out the quantum formalism as a set of 5 postulates, all of which are necessary to an orthodox or Copenhagen-like view~\cite{Jammer1974-JAMTPO-10}. These include the completeness of the state and the projection postulate. To produce something like a Copenhagen interpretation, this is supplemented by a kind of positivist objection to ascribing as real anything that cannot be directly measured, generating what Shimony~\cite{shimony1993} described as ``doing epistemology without ontology''. This all results in a similar view to Home, that it is the primacy of the quantum state that typifies the orthodox view. 
Interestingly, Stapp argues~\cite{stapp1972} that a Copenhagen viewpoint does not see the quantum state as describing the evolution of the actual system in question, only its probabilities. This seems somewhat difficult to square with the notion that the state is the most complete specification of the real situation, without some explanation of why additional information is inaccessible. 

Home~\cite{Home1997} argues that the primary characteristic of an orthodox viewpoint is that the state is framed as a complete description of a quantum system. Thus, this encapsulates both Copenhagen-like and ensembles-based theories. This latter inclusion stems from a claim that ensemble theories are internally incomplete unless they assume completeness of the quantum state (otherwise they must supply some hidden variables to assign definite values to measurable properties). Home refers to Feyeraband's encapsulation of the standard position as being that quantum systems possess no definite properties over and above those that can be derived from the state itself~\cite{feyeraband1962}. 

These points reveal a tension in descriptions of a quantum orthodoxy: a Copenhagen-like view wants to claim a ``completeness'' or ``finality'' for QM (the ``end of the road hypothesis'' as Popper termed it~\cite{Popper1982-POPQTA}). However, it also contains a strong degree of instrumentalism, as acknowledged by both Home and Jammer. This latter aspect would deny any ontological aspects to the theory. It follows then that the state is merely a prescription for the prediction of experimental outcomes. It seems clear that a follower of an allegedly Copenhagen viewpoint could pick and choose between these two, somewhat opposed, strands of thought. 

Rather than characterise this Copenhagen model, it seems more productive to instead determine what the consensus within the physics community actually is. With a particular focus on answers to questions about quantum phenomena, along the lines of those used in \cite{sivasundaram2016surveyingattitudesphysicistsconcerning}. Existing surveys of physicists tend to reveal a very fragmented community~\cite{sivasundaram2016surveyingattitudesphysicistsconcerning,survey2025}, instead we will look for a quantum mechanical orthodoxy in the many treatises that have followed from the foundational works usually analysed to define a Copenhagen interpretation. 


The target of our survey will be textbooks. The reason for this is that these works are presented for the education of a new generation of scientists. Thus, they are discussions of what we generally believe to be the `established facts' of a given field. The key question is would the `established facts' include an interpretive viewpoint? Perhaps textbooks function merely to convey a mathematical framework to their reader, with any interpretive remarks being negligible asides. This suggestion might have some validity, as a student of science can very often develop their intuition and understanding of a subject via that of the mathematics representing it. However, this is not straight-forwardly so in the realm of quantum mechanics. Evidenced by the proliferation of wildly different interpretations sharing the same (or very similar) mathematics. Thus, it is insufficient to pass on only a common mathematical core in this subject area. After all, most physics students will sketch out strikingly similar explanations of various unusual quantum results, even if their later work is done in deeply unrelated fields. They will also not merely bore a causal interlocutor with a barrage of equations. Where then would a student of quantum mechanics be exposed to an orthodoxy if not their textbooks? Perhaps in popular science media? However, one suspects that such media would itself be based on textbooks, rather than wade through the mire of competing interpretations in the active literature. An aspect of this can be seen in the fact that more recent textbooks devote more space to the interpretation of quantum phenomena (some have this as a separate chapter even). In addition, while working physicists may later deviate from the interpretations presented in reference works, we must acknowledge that their trajectory will be influenced by the initial conditions. Hence, the dominance of Copenhagen-like presentations in textbooks might lead to unorthodox interpretations which share many characteristics of this `reference orthodoxy'. An example of this would be the insistence that a quantum state does not apply merely to ensembles (as will be seen this is one of the most common assertions in textbooks). This is inherited by the vast majority of modern unorthodox interpretations (e.g. Bohmian mechanics~\cite{bohm1982}, quantum Bayesianism~\cite{Fuchs_2013}, consistent histories~\cite{sep-qm-consistent-histories}, and relational quantum mechanics~\cite{Rovelli_1996}), despite it not being necessary to the formalism~\cite{ballentine1998quantum}. Thus, even the beliefs of established researchers in quantum interpretation may be strongly correlated with the orthodoxy found in textbooks. It is therefore suggested that one cannot dismiss the interpretive remarks made in reference works on this subject. They fulfil a vital function in passing along a common understanding of quantum mechanics that is insufficiently conveyed by the mathematical formalism alone.

This work attempts an empirical, if unavoidably subjective, investigation of the quantum orthodoxy via a survey of popular and widely used textbooks on quantum mechanics, for both graduates and undergraduates. Since we are investigating interpretation (or perhaps beliefs) this cannot be wholly objective. First, one must select which books to survey, there being no means of defining a representative sample. Our sample encompasses $\nbooks$ works, with dates of first publication spanning 1930 to 2025. The sample starts with a set of works listed as ``notable'' on Wikipedia\footnote{\url{https://en.wikipedia.org/wiki/List_of_textbooks_on_classical_mechanics_and_quantum_mechanics}} and expands it via textbooks that rank among Amazon bestsellers or are written by notable authors. This is not a rigorously chosen sample, if such a thing could even be defined in this context. However, we are searching for points of consensus, so a large sample containing the most popular works should be sufficient for our purposes. Additionally, bias from sample selection is likely to be less of an issue here than in surveys of physicists themselves. 

To determine an orthodoxy empirically, we require some kind of framework of positions that might be part of the orthodoxy. Once again, this injects subjectivity into our analysis, as it is not clear how we might choose these positions. We settle on a set of 13 propositions, divided into three categories. These are selected to at least cover the major assertions of a Copenhagen-like framework and show strong overlap with surveys~\cite{sivasundaram2016surveyingattitudesphysicistsconcerning}. The first category explores how we understand the quantum state, with the propositions: ``the state is ontologically complete'', ``the state does not merely describe an ensemble of identically prepared systems'', and ``the state collapses upon measurement''. The second category revolves around the nature of particles with the propositions: ``individual particles have wave-like properties'', ``fundamental wave-like behaviour is reflected by the wave-particle duality'', ``particles cannot simultaneously possess definite values of non-commuting observables'', ``particles do not follow classical-like trajectories'', and ``there is no adequate single picture of particle or wave''. Next, we turn to the question of why quantum mechanics predicts statistics only. Here we have the propositions: ``measurement inherently disturbs quantum systems'', ``quantum systems have no property values till measured'', ``the uncertainty principle results from measurement disturbance'', and ``the limitation to prediction of probabilities results from the uncertainty principle''. Finally, we have ``other interpretations of quantum mechanics are irrelevant'', which doesn't fall into any of the categories. These will all be elaborated on in later sections of this paper. Each work is scrutinised, and a trinary value is assigned to their agreement with each proposition (an in-between value is used when agreement can be strongly inferred for consistency with other positions taken). It can then be determined to what degree the works assent to each proposition on average and, by analysing the statistics, a set of propositions that represent a consensus viewpoint can be extracted (should one exist).  

This paper is structured as follows: we first lay out the candidate propositions in section~\ref{sec:prop}, then we detail the sample of textbooks considered in section~\ref{sec:sample}, display example analyses of texts in section~\ref{sec:analysis}, and discuss the statistical properties in section~\ref{sec2}. Finally, we draw conclusions in section~\ref{sec13}.

\section{Propositions of an orthodoxy}
\label{sec:prop}
Defining all aspects of an interpretative orthodoxy is a difficult proposition. The approach taken here is to define three sub-categories of hopefully fundamental aspects of the interpretation of the quantum formalism, these being the nature of the quantum state, that of particles, and why our predictions are probabilistic. Within each category we include propositions that source works can either agree with or dissent from. The selection of these questions was designed partly apriori and partly by assertions encountered in some works. We reiterate that the choice of textbooks over research articles is made because we feel that the most distilled consensus is to be found in textbooks, which typically are formulated around what is considered ``established knowledge'' within a given field. It is entirely possible that this question set is incomplete (or contains some redundancy). 

We also note that our choices cover questions asked of working physicists in surveys~\cite{sivasundaram2016surveyingattitudesphysicistsconcerning,survey2025}. Notably, the aforementioned survey~\cite{sivasundaram2016surveyingattitudesphysicistsconcerning} also includes ``are values are defined prior to measurement?'' (``no pre-measurement values''), ``is randomness inherent in QM?'' (strong overlap with ``ontological completeness''), ``where is the electron in a hydrogen orbital?'' (``no trajectories''), ``is macroscopic superposition possible?'' (linked to ``the state doesn't just describe an ensemble''), and a question on the solution of the measurement problem is covered by the entire set of questions on why QM is probabilistic. However, we do not include questions about the role of the observer or the implications of Bell inequality violations. These latter points are highly contentious in the literature and do not receive consistent coverage in textbooks. The cases of incomplete overlap are down to the luxury of expositive works like textbooks. In that, our questions can be simpler, more numerous, and less nuanced than those in \cite{sivasundaram2016surveyingattitudesphysicistsconcerning,survey2025}, as they need not be processed in real time by the respondent.

\subsection{What is the quantum state?}
The quantum state $\psi$ stands at the heart of the quantum formalism. Thus, it is an important interpretative task to decide what this mathematical object represents in the real world. The following propositions are designed to examine some common assertions to determine whether they form part of a quantum orthodoxy.

\subsubsection{The state is ontologically complete}
This proposition indicates that there is no further information about the `real state of affairs' that can be obtained once one knows the quantum state of a given system. This stands opposed to cases like the de Broglie-Bohm interpretation, where the state is supplemented by hidden variables and a pilot wave equation. Another viewpoint that would disagree with this proposition is one that assigns only epistemic value to the state. On the other hand, an ensemble interpretation might deny this proposition, but affirm that QM is statistically complete, not because it supplements the formalism or metaphysics, rather it leaves open that QM may be a quantum analogue of statistical mechanics. 

A question arises here in how we delineate epistemic viewpoints. In this work we take ``the state contains all that can be known about the system'' as assent to ontological completeness. The ``known'' aspect might make this seem questionable. However, we focus on the ``can be known'', which implies an ontological limit on knowledge. If a work asserts that the state is a ``calculational tool only'' or that it ``only characterises our knowledge of the system'', we will deem this dissent from ontological completeness. 

\subsubsection{The state does not just describe ensembles}
Here we mean that $\psi$ describes single realisations of a given system (shorthand ``non-ensemble $\psi$''). It is not limited to the statistics of an ensemble of similarly prepared systems. This means that $\psi$ represents the current state of any single run through a given experiment. Affirming the proposition requires that individual systems can exist in indefinite/superposition states.

\subsubsection{The state collapses on measurement}
The state is capable of superposition due to the linearity of the quantum formalism. This means that the state of a system can be a linear combination of mutually exclusive states. If this state is not merely epistemic, or defined over an ensemble, we need to explain why measurements only ever return a single eigenvalue. A very common means of doing so is that the state ``collapses'' from an indefinite superposition to a single definite eigenstate upon measurement.

\subsection{The nature of particles}
It is commonly asserted that particles exhibit wave behaviour in quantum mechanics. These propositions examine different formulations of these ideas to find out if any belong to an existing orthodoxy, as well as some related questions about particles in QM.

\subsubsection{Particles are wave-like}
Individual particles exhibit wave behaviours such as diffraction and interference. The waves do not merely appear in the statistics of quantum systems. Viewpoints that might deny this proposition would be a de Broglie-Bohm interpretation~\cite{bohm1982} (where wave and particle co-exist as linked entities) or an ensemble one (where the waves are merely statistical patterns).   

\subsubsection{Wave-particle duality is necessary}
This is an assertion of the fundamental importance of wave-like behaviour, even if individual particles do not display wave behaviour. We have dual, mutually exclusive, descriptions of the same experiments. This can be viewed as a weaker form of the preceding proposition.

\subsubsection{Complementarity}
This proposition is usually specified as encompassing both the wave-particle duality and uncertainty principle. In this sense, it is a declaration that quantum systems have complementary properties that cannot be measured simultaneously. Importantly, this means there is no ``single picture'' that can be used to describe quantum systems, i.e. both wave and particle pictures are incomplete. It is this last statement that will be used to determine if a work supports complementarity. Even though it might be considered as being unnecessary, if it is given by two existing propositions, it is interesting to see if it is mentioned explicitly (as it sometimes presented as a key aspect of QM).

\subsubsection{Particles do not have trajectories}
There are no unambiguous trajectories followed by particles. These are an apparent phenomenon on the classical level only.

\subsubsection{The uncertainty principle applies to single particles}
A common joke about QM goes: Heisenberg is stopped by traffic police who ask him ``Do you know how fast you were going?''. His reply is ``No, but I know exactly where I am''. This is the essence of this proposition. That uncertainty applies to individual realisations of quantum systems. That is, if I measure the position of a particle I cannot also know its momentum simultaneously. More generally: systems cannot possess definite values of non-commuting observables simultaneously. This is opposed to a statistical understanding which recognises that the uncertainty principle is derived via standard deviations over an ensemble of measurements performed on similarly prepared systems~\cite{Popper1982-POPQTA}.

\subsection{Why are our predictions probabilistic?}
When we conduct quantum mechanical experiments we require measuring over an ensemble of similarly prepared systems in order to compare to the probabilistic predictions of the theory. If our interpretation of the quantum state is not ensemble-based or epistemic, we have to explain why this form of measurement should be the status quo. These propositions examine some explanations (and related ideas) that arise in the studied works, in order to determine which ones may belong to the orthodoxy.

\subsubsection{Measurement inherently disturbs the system}
Measurement inherently changes the state of a system while measuring it (except in the case of eigenstates). A viewpoint that denies this would note that non-interactive measurements exist in QM, or that measurement has no abstract definition.

\subsubsection{Uncertainty results from disturbance}
The source of the uncertainty principle is the fact that our measurements must disturb quantum systems. This results in a random statistical scatter in our measurements over an ensemble of similarly-prepared systems. An opposing view would be that uncertainty is present in quantum states even during Schr\"odinger evolution, so it cannot be associated with disturbance, or that the uncertainty principle can be validated on separate, similarly prepared samples (one set for $\hat{x}$ and the other for $\hat{p}$ for instance).

\subsubsection{Uncertainty leads to probabilities}
This proposition regards the uncertainty principle as a fundamental feature of nature. Particularly, this also requires single-particle uncertainty. The consequences of this is that measurements have probabilistic results, even though quantum mechanics is not formulated as fundamentally probabilistic. One might deny this if one recognises that the uncertainty principle is a consequence of quantum rules, rather than a postulate, or if one notes the uncertainty principle is formulated in terms of statistical observables only.

\subsubsection{There are no pre-measurement values}
Except for those observable eigenstates, quantum systems do not have definite properties that we uncover via measurement. Instead, the value comes to exist at the point of measurement. A primary signature of this is how the Born rule is described. If a text states that $\vert \psi (x) \vert^2$ is the probability of \textit{finding} the system with property value $x$, then it affirms this proposition. However, this alone is not enough, we only infer assent if there is other supportive text as well. A dissenting text would instead state the Born rule as giving the probability of a system \textit{having} property value $x$. This appears to be a rather fine point of semantics. However, almost all the surveyed texts use precisely the same wording: finding, not having. The ones that don't use ``finding'', seem to do so deliberately, and it correlates with more direct assertions that pre-measurement values exist.

\subsection{Other propositions}
These don't fit into the above categories.

\subsubsection{There is only the orthodoxy}
Only the orthodox presentation of quantum mechanics is worthwhile or serious. An affirming text either mentions no other interpretations or dismisses them all as unconvincing or disproven. A textbook that notes it is presenting the orthodox/conventional/standard approach but admits this isn't definitive will be taken as dissenting. 


\section{Sample of studied works}
\label{sec:sample}
To perform this study we examine the $\nbooks$ popular textbooks listed in Table~\ref{tab:sample}. These range in date of first publication from 1930 to 2025. This large time span is of utility, as it allows us to determine whether there is any evolution within the consensus view of quantum mechanics. Additionally, we consider subdivision of these works into graduate ($N=\ngbooks$) and undergraduate ($N=\nugbooks$), as determined from the statements of the target audience in the work itself. These works were chosen to encompass the Wikipedia list on `Notable quantum mechanics textbooks'\footnote{\url{https://en.wikipedia.org/wiki/List_of_textbooks_on_classical_mechanics_and_quantum_mechanics}} ($N=\nwikibooks$) as well as to include as large a sample as feasible. To detect variation over time we also consider three subsamples: works pre-1980 ($N=\noldbooks$), post-1980 ($N=\nmodernbooks$), and post-2000 ($N=\nmillbooks$). Note that Omn\'es work is included, even though it is a textbook that advocates against the orthodoxy in favour of ``consistent histories'', as it provides its own description of the orthodox positions which we use for our study. In the case of \cite{rae2002}, this thoughtfully discusses some foundational issues without expressing a direct opinion, here we substitute what the author characterises as the ``conventional'' approach. We will also make reference to a set of 4 works whom we do not include in any of the samples: \cite{Home1997,Hannabuss:1999fa,nielsen2010,ballentine1998quantum}. The first is excluded as it is a discussion of interpretations, but will provide a useful point of comparison. The second two are analysed out of curiosity, but are not targeted squarely as part of a physics syllabus. The final one will be analysed and found to be entirely outside the orthodoxy on almost every point of contention (so it is treated as an outlier).  
\begin{table}[htbp]
	\centering
	\caption{Sample of studied works. Note that the column `Year' is that of first publication. `Level' indicates undergraduate (U) or graduate (G). `Wiki' is Y if the book is on the wikipedia list for notable textbooks in quantum mechanics. The four books at the bottom are surveyed but not included in various samples.}
	\label{tab:sample}
		\begin{tabular}{|l|l|l|l|l|l|}
			\hline
			Title & Author & Year & Ref & Level & Wiki \\
			\hline
			Lecture notes on quantum mechanics & D. Tong & 2025 & \cite{tong2025} & U & N \\
			Basic quantum mechanics & K. Tamvakis & 2019 & \cite{tamvakis2019} & U & N \\
			A first introduction to quantum physics & P. Kok & 2018 & \cite{kok2023} & U & N \\
			Lectures on quantum mechanics & S. Weinberg & 2013 & \cite{weinberg2013} & G & N \\
			The physics of quantum mechanics & J. Binney \& D. Skinner & 2013 & \cite{binney2013physics} & U & Y \\
			Quantum concepts in physics & M. Longair & 2013 & \cite{longair_2013} & G & N \\
			Quantum mechanics & D. McIntyre et al & 2012 & \cite{mcintyre2012quantum} & U & Y \\
			Quantum mechanics: theory and experiment & M. Beck & 2012 & \cite{beck2012} & U & N \\
			Quantum mechanics: a new introduction & K. Konishi \& G. Paffuti & 2009 & \cite{Konishi:2009qva} & U & N \\
			Quantum mechanics for scientists and engineers & D. Miller & 2008 & \cite{miller2010} & U & N\\
			Quantum mechanics: a textbook for undergraduates & M. Jain & 2007 & \cite{jain2017} & U & N \\
			Introduction to quantum mechanics & H. M\"uller-Kirsten & 2006 &\cite{muller-kisten2006} & U & Y \\
			Quantum physics & M. Le Bellac & 2006 & \cite{bellac2011quantum} & G & N \\
			Lectures on quantum mechanics & J. Basdevant & 2005 & \cite{basdevant2023} & G & N \\
			Introduction to quantum mechanics & A. Philips & 2003 & \cite{phillips2013introduction} & U & N \\
			Quantum Mechanics: Concepts and Applications & N. Zettili & 2001 & \cite{zettili2001} & U & Y \\
			Elements of quantum mechanics & M. Fayer & 2001 & \cite{fayer2001} & U & N \\
			Understanding quantum mechanics & R. Omn\'es & 1999 & \cite{Omnes1999-OMNUQM} & U & N \\
			Particles behave like waves & T. Moore & 1998 & \cite{moore1998} & U & N \\
			Introduction to quantum mechanics & D. Griffiths & 1995 & \cite{Griffiths2004Introduction} & U & Y  \\
			Quantum theory & A. Peres & 1993 & \cite{peres1993} & G & Y \\
			Quantum mechanics & F. Mandl & 1992 & \cite{mandl2013quantum} & U & N  \\
			A modern approach to quantum mechanics & J. Townsend & 1992 & \cite{townsend2012modern} & U & Y  \\
			Quantum mechanics: an introduction & W. Greiner & 1989 &\cite{greiner1989}& U & N  \\
			Quantum mechanics & A. Rae & 1986 & \cite{rae2002} & U & N \\
			Modern quantum mechanics & J. Sakurai & 1985 &\cite{Sakurai:1167961} & G & Y  \\
			Principles of Quantum mechanics & R. Shankar & 1980 & \cite{Shankar:102017} & G & Y \\
			An introduction to quantum physics & A. French \& E. Taylor & 1978 & \cite{french1978} & U & Y  \\
			Quantum mechanics & C. Cohen-Tannoudji & 1977 &\cite{Cohen-Tannoudji:101367} & G & Y \\
			Quantum physics of atoms, molecules, solids, nuclei  & R. Eisberg \& R. Resnick & 1974 & \cite{eisberg1974} & U & Y \\
			and particles & & &  & & \\
			Quantum physics & S. Gasiorowicz & 1974 &\cite{gasiorowicz2007quantum} & U & Y \\
			Feynman lectures on physics volume 3 & R. Feynman & 1965 & \cite{feynman1977feynman} & U & Y \\
			Quantum mechanics & A. Davydov & 1965 & \cite{davydov1965} & G & Y \\
			Elementary quantum mechanics & P. Fong & 1962 & \cite{fong2013elementary} & U & N \\
			Quantum mechanics & E. Merzbacher & 1961 & \cite{merzbacher1998} & G & N\\
			Introduction to quantum mechanics & R. Dicke \& J. Wittke & 1960 & \cite{dicke1960} & G & N \\
			Quantum mechanics & A. Messiah & 1959 &\cite{messiah1999quantum} & G & N \\
			Quantum mechanics: non-relativistic theory & L. Landau \& L. Lifshitz & 1958 & \cite{Landau1981Quantum} & G & Y \\
			Mathematical foundations of quantum mechanics & J. von Neumann & 1955 & \cite{vonneumann1955} & G & Y \\
			Quantum theory & D. Bohm & 1951 & \cite{bohm1951} & G & N  \\
			Quantum mechanics & L. Schiff & 1949 &\cite{schiff1955} & G & N \\
			The principles of quantum mechanics & P. Dirac & 1930 & \cite{Dirac1930-DIRTPO} & G & Y \\
			\hline
			Quantum mechanics: a modern development & L. Ballentine & 1998 & \cite{ballentine1998quantum} & G & N \\
			Quantum computation and quantum information & M. Nielsen \& I. Chuang & 2000 & \cite{nielsen2010} & G & N \\
			An introduction to quantum theory & K. Hannabuss & 1997 &\cite{Hannabuss:1999fa} & G & N \\
			Conceptual foundations of quantum physics & D. Home & 1997 & \cite{Home1997} & - & N \\
			\hline
		\end{tabular}
\end{table}

\section{Analysis from example texts}\label{sec:analysis}
To illustrate the analysis process we will consider several texts as examples. This allows the reader to determine that we are carefully evidencing degrees of assent or dissent in each case. Note that the sections below leave out propositions that are unaddressed by the given text.

\subsection{Binney \& Skinner}
We begin with \textit{The physics of quantum mechanics} by Binney \& Skinner~\cite{binney2013physics}. For ``measurement disturbs systems'' we find initial evidence on page 3: ``Moreover, the act of measurement inevitably disturbs the atom, and leaves it in a different state from the one it was in before we made the measurement'' (note that this is not in the context of the physics of a particular type of measurement). This is reinforced on page 130 in a discussion of the collapse hypothesis: ``The Copenhagen interpretation does, however, contain a crucial insight into measurement by stressing that any measurement physically disturbs the system, so the system's state after a measurement has been made is different from what it was earlier''. For ``uncertainty linked to disturbance'' we find evidence for endorsement on page 3 (``We rarely know the exact state that our measuring instrument is in before we bring it into contact with the system we have measured, so the result of the measurement of the atom would be uncertain even if we knew the precise state that the atom was in before we measured it, which of course we do not''), page 28 (a discussion of the uncertainty principle in terms of measurement uncertainty) which is subsequently reinforced on pages 130 and 131. For single particle uncertainty we find on page 28 ``the position of a particle that has well-defined momentum is maximally uncertain'' as evidence for adherence to this position. For ``particles are wave-like'' we find a discussion opposing this on page 7, particularly ``we cannot consistently infer from this correspondence that particles are manifestations of waves''. For ``collapse'', the work opposes this idea, explicitly stating collapse is not physical on page 15 and that the hypothesis is false on page 130. For ``particle-wave duality'' the authors never mention the concept explicitly but already oppose ``single particles are wave-like'' outside of quantum field theory contexts (page 7). For ``non-ensemble $\psi$'' we find on page 14 ``in quantum mechanics an individual photon has an amplitude to be linearly polarised in any chosen direction and an amplitude to be circularly polarised in a given sense'', so states correspond to individual system realisations. This is reinforced on page 71 where they state ``In quantum mechanics, complete information about any system is contained in its ket $\ket{\psi}$''. For ``probability from uncertainty'' the authors dissent, as we have seen that they attribute probability to the uncertain state of the measuring apparatus and its subsequent disturbing of the system. For ``no pre-measurement values'', their view on pages 130--131 seems to be that systems are disturbed by measurement, changing the values found rather than creating them altogether (the Born rule is stated as a probability of finding a value). We have already seen that the authors view the state as ontologically complete, an ``extended'' theory is possible, but it merely accounts for the actual physics of measurement (page 130 and 131). There is no mention of other viable interpretations. 

\subsection{Tamvakis}
Now we consider \textit{Basic quantum mechanics} by Tamvakis~\cite{tamvakis2019}. On page 86 it is outlined, as part of the axioms of quantum mechanics, that the state changes during measurement indicating support for ``measurement disturbs systems''. On page 5 uncertainty is described in terms of single particle measurement precision or ``if we make a measurement on any object and determine its position along a given direction with uncertainty $\Delta x$, its momentum along this direction will necessarily be known with uncertainty no less than $\frac{h}{4\pi \Delta x}$'', an indication of ``single particle uncertainty''. On page 6, the de Broglie relation is discussed as the wave aspect of a single particle, implying individual particles are wave-like. Pages 86 and 87 discuss a non-unitary evolution during measurement and call this collapse, this re-iterated on page 442. On page 5: ``the particle aspects of a system are complementary to its wave aspects'' directly supports complementarity. On page 7 it is stated that ``the wave function of the particle characterises the system fully at each point in time''. This indicates ontological completeness and a single realisation/particle association also supported by the previous quote. Page 5 describes the uncertainty principle in terms of knowledge of $x$ and $p_x$ ``thus, viewed from a classical viewpoint, in a microscopic system, complete knowledge of $x$ and $p_x$ is impossible'', we infer this is indicating that probabilities result from incomplete knowledge implied by the uncertainty principle. For ``no pre-measurement values'', we note that it is stated on page 86 that a superposition represents the pre-measurement state, since this cannot have a definite property value we must conclude that definite property values are created, or at least changed, during measurement. This will satisfy an assent to the proposition  as the author describes measurement as ``acausal'' on page 87, so the initial state is irrelevant (the Born rule is stated as a probability of finding a value). There is no mention of other viable interpretations.

\subsection{Zettili}
Next we consider \textit{Quantum Mechanics: Concepts and Applications} by Zettili~\cite{zettili2001}. On page 24, ``measurement interferes with the state of microscopic objects'' and ``it is the effects of interference by the equipment on the system which is the essence of quantum mechanics'' on page 165 demonstrate support for measurement disturbance. On page 169 uncertainty is described as a result of measurement ``perturbing'' the system. ``Single particle uncertainty'' is affirmed on page 28 by ``If the $x$ component of the momentum of a particle is measured with an uncertainty $\Delta p_x$, then its $x$ position cannot, at the same time, be measured more accurately than $\Delta x = \frac{\hbar}{2 \Delta p_x}$''. Support for single particles being wave-like is given on page 18: ``each material particle behaves as a group of waves''. Pages 24 and 25 indicate ``we cannot follow a microscopic particle along its motion, nor can we determine its path'', while 51 discusses wave packet spreading only meaning the ``position cannot be known exactly''. These are taken to indicate trajectories exist but are unknown. On page 158 collapse is included in the postulates of quantum mechanics. On page 26 ``Microscopic systems, therefore, are neither pure particle nor pure waves, they are both'', as well as direct invocations on the same page, suffices for duality and complementarity. For ontological completeness, page 158 asserts ``the state contains all the needed information about the system''. On page 30 we find ``In quantum mechanics the state (or one of the states) of a particle is described by a wavefunction $\psi(r,t)$ corresponding to the de Broglie wave of this particle; so $\psi(r,t)$ describes the
wave properties of a particle'' indicates that the state applies to single instances. We combine previous statements from 165 and 169 to conclude that probability is introduced by measurement disturbance (uncertainty). Based on the axioms from 158, where the Born rule is stated as a probability of finding a value, as well as the state being changed to an eigenstate on measurement, we see assent to ``no pre-measurement values''. There is no mention of other viable interpretations. 

\subsection{McIntyre, Manogue, \& Tate}
We now turn to \textit{Quantum mechanics} by McIntyre, Manogue, and Tate~\cite{mcintyre2012quantum}. On page 8 we find that measurements disturbing systems is a ``key feature'' of quantum mechanics. On page 59, in the context of the uncertainty principle, it is stated that ``the lack of ability to measure all spin components simultaneously implies that the spin does not really point in a given direction''. This indicates the uncertainty principle is being applied to individual particles. Combining this with similar discussion on 208 (about position and momentum) implies a link between uncertainty and disturbance. A confusing discussion on 207 about the double-slit seems to suggest both that a wave picture is just for finding particle probabilities but also that the waves have to actually interfere (passing through two slits). The same discussion refers the reader to Feynman and Cohen-Tannoudji which seems to indicate the ultimate outcome is individual particles are wave-like too. Page 192 indicates we cannot know a particle's trajectory but does not deny its existence. Page 48 includes collapse in the postulates. Wave-particle duality is explicitly invoked as a correct description of quantum systems on pages 121 and 207. Complementarity in terms of position and momentum is discussed on pages 192 and broadened to particle and wave properties in 207 and 208. ``The state of quantum mechanical system, including all the information you can know about it, is represented mathematically by a normalised ket $\ket{\psi}$'' covers ontological completeness. On page 14 the authors describe the Born rule as yielding the probability to measure a given value. On page 19, they say superpositions are not probabilistic mixtures. These last two, combined with page 8's description of superposed states as having indefinite values indicates assent to ``no pre-measurement values''. The authors indicate that matters of interpretation are irrelevant, and advise the reader to ``shut up and calculate'' on page 168.      

\subsection{Konishi \& Paffuti}
Finally, we review \textit{Quantum mechanics: a new introduction} by Konishi \& Paffuti~\cite{Konishi:2009qva}. On page 29 the authors support the notion that all measurement causes disturbance by stating that ``the disturbance caused by the process of measurement on the microscopic system under examination cannot be controlled beyond a certain limit.'' The unqualified illustration of the uncertainty principle via ``Heisenberg microscopes'' on page 28 indicates the authors link uncertainty to measurement disturbance. On page 26 we find ``the classical concept of a particle trajectory no longer makes sense. An electron cannot simultaneously have a definite position and momentum.'' This demonstrates both that the uncertainty principle is applied to individual particles and that the authors reject trajectories for quantum objects. This comes with a caveat on page 29 that uncertainty principle is not itself fundamental, but derives from the ``dynamical properties of the physical processes involved in the measurement.'' On page 6 the authors indicate that single particles are wave-like by asserting that wave-particle duality ``concerns the properties of an individual electron'' and not a population. This, combined with page 28 highlighting the importance of wave-particle duality to uncertainty relations, supports a more general duality thesis too. On pages 31 and 32 we find that the authors believe state reduction is ``necessary for the theory to make sense at all.'' This is furthered on 518 and 519 where they argue collapse is experimentally confirmed. On page 25 the authors describe a superposed polarisation state as ``the state of a single photon''. Additionally, on page 23 the authors ascribe ``physical properties'' to superposition states that are ``somewhere between'' those of the constituents. These indicate that quantum states do not apply to ensembles but to single realisations of systems. On page 29 the authors agree that probability derives from the uncertainty principle ``the probabilistic nature of quantum mechanical prediction for the results of measurement is an inevitable consequence of Heisenberg's uncertainty principle.'' Although the discussion on 518 and 519 is ambiguous to whether values are being revealed or created (the authors use ``indefinite'' properties of a superposition here to mean ``unpredictable''), we must infer that there are no pre-measurement values following the assertion, on page 23, that superposition properties are intermediate to definite states. This is reinforced by phrasing the Born rule in terms of ``finding property values'' on page 31. On page 21 the authors state that ``knowledge of the wave function amounts to complete knowledge of the state'' which, combined with prior assertions fully supports ontological completeness. There is no mention of other viable interpretations. 

\section{Results and discussion}\label{sec2}

In Table~\ref{tab:data} we display all the data used for this study (pages referenced are listed in Table~\ref{tab:pages}). Each work's response to the propositions is characterised by a degree of agreement. This is kept as simple as possible with three principle values: $1$ indicates agreement, $-1$ dissent, and $0$ as not mentioned. We expand this via either of $\pm 0.5$ to mean `inferred for logical consistency'. Inference of agreement/dissent is only made if the case is strong (in order to allow for a conservative analysis). We proceed to study the responses by observing the frequency of agreement. Note that this computation counts $\pm 0.5$ responses as half a count when determining frequencies. Our raw data keeps the page references separate for each proposition (available on request).

\begin{sidewaystable}[htbp]
	\caption{Table of all the studied references. The column headings indicate propositions (abbreviated for space). A $1$ indicates agreement, $0$ dissent, and $0.5$ means `inferred for logical consistency'. All the pages used to create this table are listed in the final column. MDS = Measurement Disturbs System. ULD = Uncertainty Linked to Disturbance. 1PU = Single Particle Uncertainty. 1PW = Single Particle Wave-like. NT = No Trajectories. Coll = Collapse. WPD = Wave-Particle Duality. Com = Complementarity. SI$\psi$= Single Instance $\psi$. PFU = Probability From Uncertainty. OC = Ontologically Complete. NPMV = ``no pre-measurement value''. OO = ``orthodoxy only''.}\label{tab:data}
		\begin{tabular}{|lcccccccccccccc|}
			\hline
			Author & Year & MDS & ULD & 1PU & 1PW & NT & Coll & WPD & Com & 1P$\psi$ & PFU & OC & NPMV & OO \\
			\hline
			Tong & 2025 & 1 & 1 & 1 & 1 & 1 & 1 & -1 & 0 & 1 & 1 & 1 & 1 & 1 \\
			Tamvakis & 2019 & 1 & 0 & 1 & 1 & 0 & 1 & 1 & 1 & 1 & 1 & 1 & 1 & 1 \\
			Kok & 2018 & 1 & 1 & 1 & 1 & 1 & 1 & 1 & 0 & 1 & 0 & 0 & 1 & 0.5 \\
			Weinberg & 2013 & 0 & 0 & 0.5 & -1 & 0 & -1 & 0 & 0 & 1 & 0 & -1 & 1 & -1 \\
			Binney and Skinner & 2013 & 1 & 1 & 1 & -1 & 0 & -1 & -1 & 0 & 1 & -1 & 1 & -1 & 1 \\
			Longair & 2013 & 1 & 1 & 1 & 1 & 0.5 & 1 & 1 & 1 & 1 & 1 & 1 & 1 & 1 \\
			McIntyre, Manogue, and Tate & 2012 & 1 & 0.5 & 1 & 1 & -1 & 1 & 1 & 1 & 1 & 0 & 1 & 1 & 1 \\
			Beck & 2012 & 1 & 0 & 1 & 1 & 0 & 1 & 0 & 1 & 1 & 0 & 1 & 1 & 1 \\
			Konishi and Pafutti & 2009 & 1 & 1 & 1 & 1 & 1 & 1 & 1 & 0 & 1 & 1 & 1 & 1 & 1 \\
			Miller & 2008 & 1 & 0 & 1 & 1 & 0 & 1 & 1 & 0 & 1 & 0 & 1 & 1 & -1 \\
			Jain & 2007 & 1 & 1 & 1 & 1 & 0 & 1 & 1 & 1 & 1 & 0 & 1 & 1 & 1 \\
			Muller-Kirsten & 2006 & 1 & 0 & 1 & 1 & 0 & 0 & 1 & 1 & 1 & 1 & 1 & 1 & 1 \\
			Le Bellac & 2006 & 1 & 0 & 1 & 1 & 0 & 1 & 1 & 0 & 1 & 0 & 1 & 1 & 1 \\
			Basdevant & 2005 & 1 & -1 & 1 & 1 & 1 & 1 & 1 & 1 & 1 & 0 & 1 & 1 & 1 \\
			Phillips & 2003 & 1 & 1 & 1 & 1 & 1 & 1 & 1 & 0 & 1 & 0 & -1 & 1 & 1 \\
			Zettili & 2001 & 1 & 1 & 1 & 1 & -0.5 & 1 & 1 & 1 & 1 & 0.5 & 1 & 1 & 1 \\
			Fayer & 2001 & 1 & 1 & 1 & 1 & 1 & 1 & 1 & 0 & 1 & 0 & 1 & 1 & 1 \\
			Omnes & 1999 & 1 & 1 & 1 & 1 & 1 & 1 & 1 & 1 & 1 & 1 & 1 & 1 & -1 \\
			Moore  & 1998 & 1 & 0 & -1 & 1 & 0.5 & 1 & 1 & 1 & 1 & 0 & -1 & 1 & 1 \\
			Griffiths & 1995 & 1 & 1 & -1 & 0 & 0.5 & 1 & 1 & 0 & 1 & -1 & 1 & 1 & 0.5 \\
			Peres & 1993 & -1 & 0 & -1 & 1 & -1 & 0 & 1 & 1 & -1 & -1 & 1 & 1 & 0.5 \\
			Mandl & 1992 & 1 & 1 & 0.5 & 1 & 0 & 0 & 0 & 0 & 0.5 & 0 & 1 & -1 & 1 \\
			Townsend & 1992 & 1 & 1 & 1 & 1 & 0 & 1 & 0 & 0 & 1 & 0 & 1 & 1 & 1 \\
			Greiner & 1989 & 1 & 0 & 1 & 1 & 1 & 1 & 1 & 0.5 & 1 & 0 & 1 & -1 & -1 \\
			Rae & 1986 & 0.5 & -1 & -1 & 1 & 1 & 1 & 1 & 1 & 1 & 0 & 1 & 1 & -1 \\
			Sakurai & 1985 & 1 & 1 & -1 & -1 & 0 & 1 & 0 & 0 & 1 & 0 & 1 & 1 & 0.5 \\
			Shankar & 1980 & 1 & 1 & 1 & 1 & 1 & 1 & 1 & 0 & 1 & 0 & 1 & 1 & 1 \\
			French and Taylor & 1978 & 1 & 1 & 1 & 1 & 0.5 & 1 & 1 & 0 & 1 & 0 & 0 & 1 & 1 \\
			Cohen-Tannoudji & 1977 & 1 & 1 & 1 & 1 & 1 & 1 & 1 & 1 & 1 & 0 & 1 & -1 & 1 \\
			Eisberg and Resnick & 1974 & 1 & 1 & 1 & 1 & 0 & 0 & 1 & 1 & 1 & 1 & 0 & 1 & 1 \\
			Gasiorowicz & 1974 & 1 & 1 & 1 & 1 & 0 & 0.5 & 1 & 0 & 1 & 0 & 1 & 1 & 1 \\
			Feynman & 1965 & 1 & 1 & 1 & 1 & 1 & 1 & 1 & 1 & 1 & 1 & 1 & 1 & 1 \\
			Davydov & 1965 & 1 & 0 & 1 & 1 & 1 & 1 & 0 & 1 & 1 & 0 & 1 & 1 & 1 \\
			Fong & 1962 & 1 & 1 & 1 & 1 & 0 & 1 & 1 & 1 & 1 & 1 & 1 & 1 & -1 \\
			Merzbacher & 1961 & 0 & 0 & 1 & 1 & 1 & -1 & 1 & 1 & 1 & 0 & 1 & 1 & 1 \\
			Dicke and Wittke & 1960 & 1 & 1 & 1 & 1 & 0 & 1 & 1 & 1 & 1 & 0 & 1 & 1 & 1 \\
			Messiah & 1959 & 1 & 1 & 1 & 1 & 1 & 1 & 1 & 1 & 1 & 0 & 1 & 1 & 1 \\
			Landau and Lifshitz & 1958 & 1 & 1 & 1 & 1 & 1 & 1 & 0 & 0 & 1 & 1 & 1 & 1 & 1 \\
			von Neumann & 1955 & 1 & 0 & 1 & 0 & 1 & 1 & -1 & 0 & 1 & 1 & 1 & 1 & 1 \\
			Bohm & 1951 & 1 & 1 & 1 & 1 & 1 & 1 & 1 & 1 & 1 & 1 & 1 & 1 & 1 \\
			Schiff & 1949 & 1 & 1 & 1 & 1 & 1 & 0 & 1 & 1 & 1 & 0 & 1 & 1 & 1 \\
			Dirac & 1930 & 1 & 1 & 1 & 1 & 0 & 1 & 1 & 0 & 1 & 0 & 1 & 1 & 1 \\
			Ballentine & 1998 & -1 & -1 & -1 & -1 & -1 & -1 & -1 & 0 & -1 & -1 & -1 & -1 & -1 \\
			Nielsen and Chuang & 2000 & 1 & 0 & -1 & 0 & 0 & 1 & 0 & 0 & 1 & 0 & 1 & 1 & 1 \\
			Hannabuss & 1997 & 1 & 0 & 0.5 & 1 & 0 & 1 & 1 & 0 & 1 & 0 & -1 & -1 & 1 \\
			Home & 1997 & 1 & 0 & -1 & 1 & 1 & 0 & 1 & 0 & 1 & 0 & 1 & 1 & -1 \\
			\hline
	\end{tabular}
\end{sidewaystable}

\begin{table}[htbp]
	\caption{Table of the pages used in studied references.}\label{tab:pages}
		\begin{tabular}{|lll|}
			\hline
			Author & Year & Pages \\
			\hline
			Tong & 2025 & 10,11,20,21,25,29,97,100,108,123,124,574\\
			Tamvakis & 2019 & 5,6,7,86,87,442\\
			Kok & 2018 & 6,58,182,186,206,211,225,226,264\\
			Weinberg & 2013 & 21,25,26,82,83,88,95,97\\
			Binney and Skinner & 2013 & 3,7,8,14,15,28,71,130,131\\
			Longair & 2013 & 178,183,346,359,360,363,367\\
			McIntyre, Manogue, and Tate & 2012 & 7,8,14,19,48,59,121,167,168,192,207,208\\
			Beck & 2012 & 62,91,92,93,101,129,218\\
			Konishi and Pafutti & 2009 & 6,21,23,25,26,28,29,518,519\\
			Miller & 2008 & 6,9,74,83,84,425,449,450\\
			Jain & 2007 & 61,77,83,86,89,209,210,214\\
			Muller-Kirsten & 2006 & 9,13,17,19,20,77,169\\
			Le Bellac & 2006 & 22,24,97,101,102,104,179,185\\
			Basdevant & 2005 & 3,17,20,35,43,47,73,79,171,420\\
			Phillips & 2003 & 6,11,13,16,21,27,50\\
			Zettili & 2001 & 18,24,25,26,28,30,51,158,165,169\\
			Fayer & 2001 & 1,2,3,5,32,33,42,55\\
			Omnes & 1999 & 31,45,47,49,53,91\\
			Moore  & 1998 & 69,77,81,106,107,108,113,135\\
			Griffiths & 1995 & 1,2,3,4,5,19,112,420,431\\
			Peres & 1993 & 5,11,14,20,24,25,26,27,93\\
			Mandl & 1992 & 3,13,50,83\\
			Townsend & 1992 & 6,8,10,13,16,25,165,211\\
			Greiner & 1989 & 29,34,38,51,53,259,469,473\\
			Rae & 1986 & 6,8,10,61,73,79,261,273,287\\
			Sakurai & 1985 & 6,11,23,24,101\\
			Shankar & 1980 & 111,112,113,115,116,129,140\\
			French and Taylor & 1978 & 72,76,81,85,92,93,120,243,317,328\\
			Cohen-Tannoudji & 1977 & 14,19,21,27,28,221,252,253,279\\
			Eisberg and Resnick & 1974 & 64,65,68,70,73,74,76,87,88,147\\
			Gasiorowicz & 1974 & 13,24,28,32,94,318,323,327,328\\
			Feynman & 1965 & 1-10,1-11,1-6,1-7,1-9,2-1,2-10,2-2,2-3,2-8\\
			Davydov & 1965 & 2,3,4,6,33,37,40,51\\
			Fong & 1962 & 28,29,32,50,51,52,53,114,116,123,124\\
			Merzbacher & 1961 & 3,4,6,9,17,18,28,408\\
			Dicke and Wittke & 1960 & 14,15,16,24,25,27,40,74,91,110,115,120\\
			Messiah & 1959 & 20,21,46,49,116,117,119,140,143,144,153,157,162\\
			Landau and Lifshitz & 1958 & 2,3,4,5,6,24,31,47,57,157\\
			von Neumann & 1955 & 4,210,214,216,247,326,403\\
			Bohm & 1951 & 59,73,80,92,99,104,117,120,145,160\\
			Schiff & 1949 & 3,6,7,8,13,16,21\\
			Dirac & 1930 & 4,7,8,9,16,34,46,52,98\\
			Ballentine & 1998 & 100,133-136,138,226,227,233,236,244,247,45,47,55\\
			Nielsen and Chuang & 2000 & 13,15,80,81,88,89,96\\
			Hannabuss & 1997 & 3,11,77,86,173,177\\
			Home & 1997 & 14,16,18,19,22,38,52\\
			\hline
	\end{tabular}
\end{table}

\begin{figure}[ht!]
	\centering
	\resizebox{0.95\hsize}{!}{\includegraphics{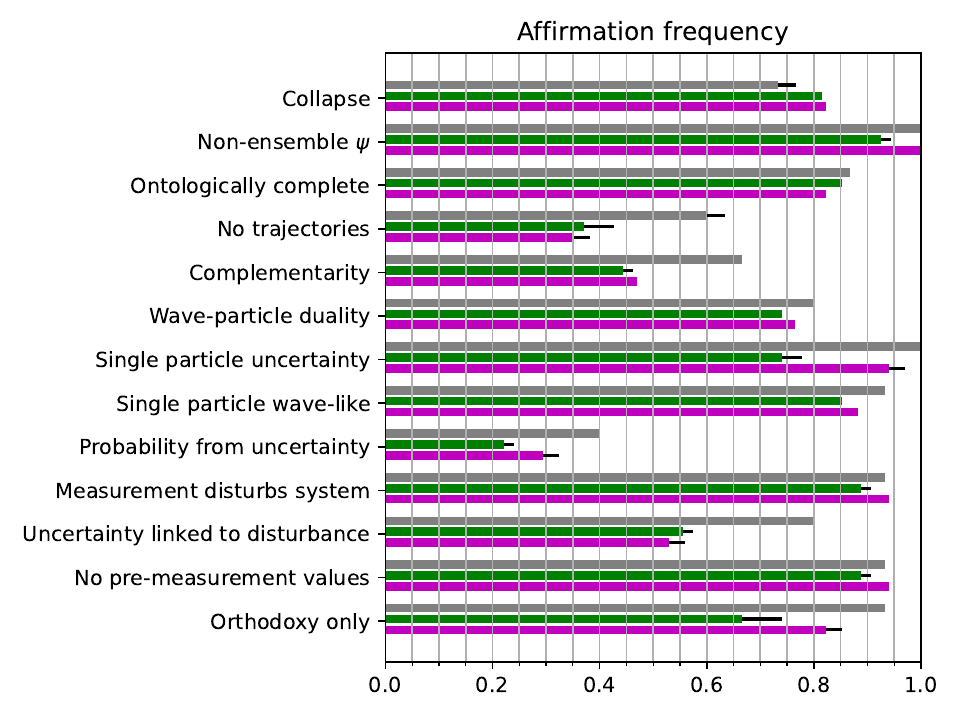}}
	\caption{Affirmation frequency for responses to each question as a function of time. The bars represent different samples: pre-1980 (grey), post-1980 (green), and post-2000 (magenta). The error bars reflect the effect of including values inferred for logical consistency.} 
	\label{fig:mode-time}
\end{figure}

\begin{figure}[ht!]
	\centering
	\resizebox{0.95\hsize}{!}{\includegraphics{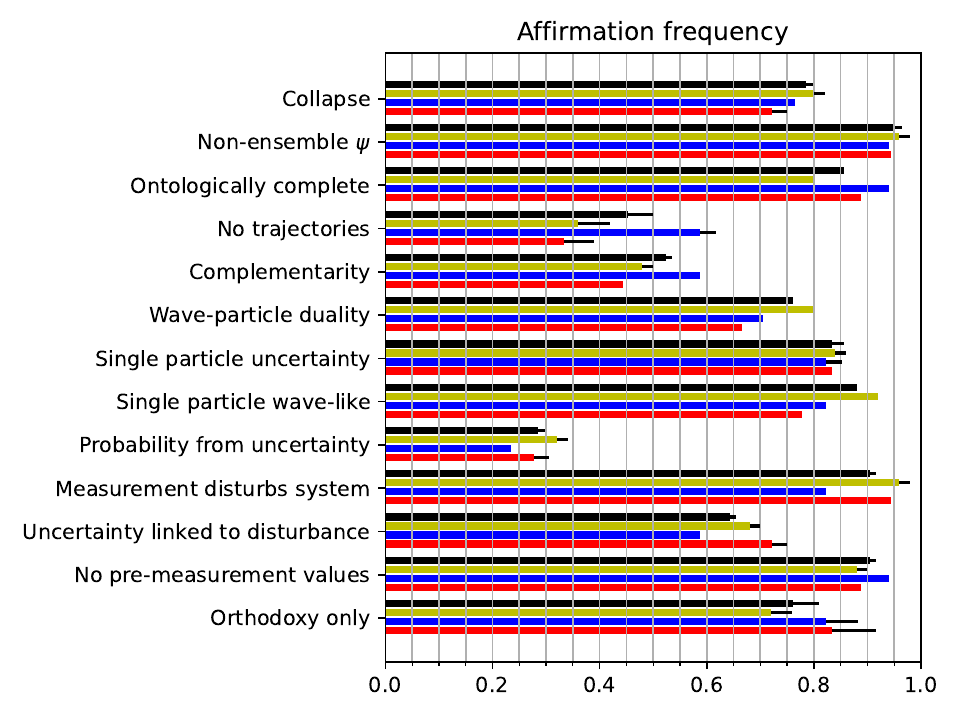}}
	\caption{Affirmation frequency for responses to each question for various subsamples. The bars represent different samples: all (black), undergraduate (yellow), graduate (blue), Wikipedia notable (red). The error bars reflect the effect of including values inferred for logical consistency.} 
	\label{fig:mode-other}
\end{figure}

\begin{figure}[ht!]
	\centering
	\resizebox{0.95\hsize}{!}{\includegraphics{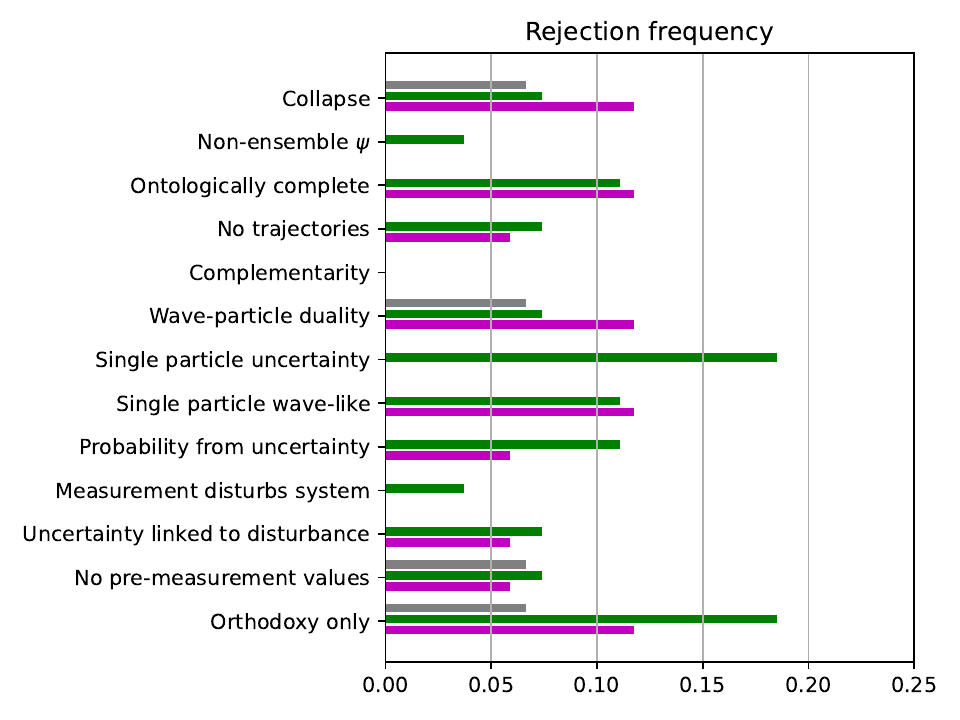}}
	\caption{Rejection frequency for responses to each question as a function of time. The bars represent different samples: pre-1980 (grey), post-1980 (green), and post-2000 (magenta). The error bars reflect the effect of including values inferred for logical consistency.} 
	\label{fig:reject-time}
\end{figure}

\begin{figure}[ht!]
	\centering
	\resizebox{0.95\hsize}{!}{\includegraphics{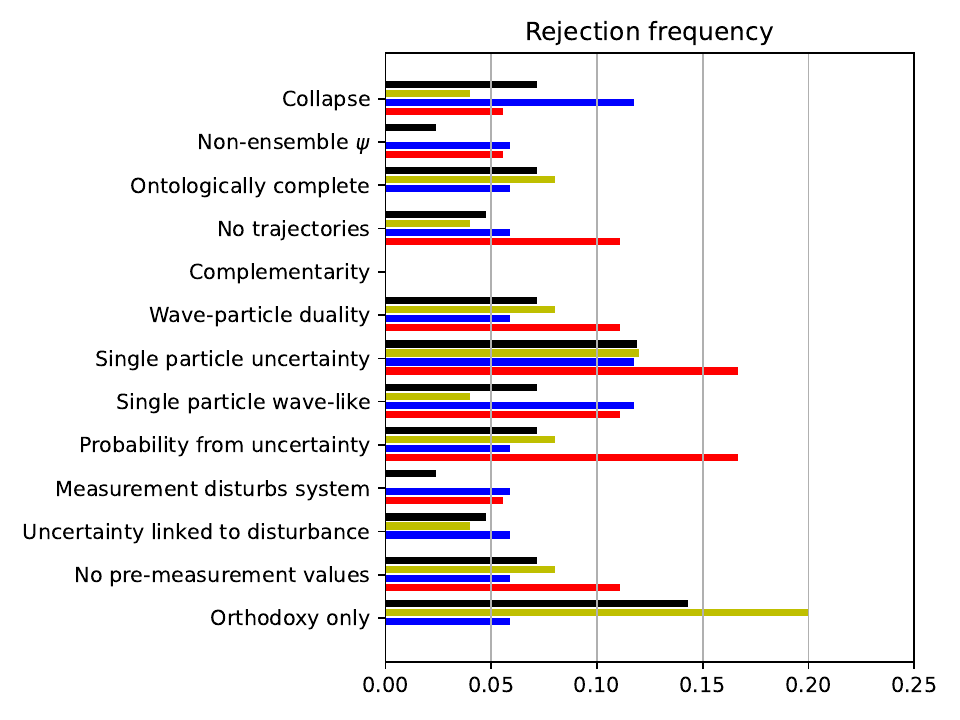}}
	\caption{Rejection frequency for responses to each question for various subsamples. The bars represent different samples: all (black), undergraduate (yellow), graduate (blue), and Wikipedia notable (red). The error bars reflect the effect of including values inferred for logical consistency.} 
	\label{fig:reject-other}
\end{figure}

In Figure~\ref{fig:mode-time} we display the agreement frequencies as a function of time via 3 subsamples: pre-1980 works (gray), post-1980 works (green), and post-2000 works (magenta). We also display the following 4 samples in Fig.~\ref{fig:mode-other}: the full sample (black), undergraduate books (yellow), graduate texts (blue), and works on the `Notable books' list from Wikipedia (red). We supplement this with rejection frequencies in Figs.~\ref{fig:reject-time} and \ref{fig:reject-other}. It would be ideal to include the significance of the consensus in each case. However, there is no obvious statistical prescription for doing so, as responses to the propositions are not generated by random processes in any sense. We note however, that many positions here achieve agreement frequencies above 80\%. Given a sample of $\nbooks$ works, this is sign of substantial consensus. 

We begin by discussing propositions related to the nature of the quantum state. Interestingly, we see that not all the books regard $\psi$ as ontologically complete. However, the support for this proposition is robust with assent frequency $\sim 0.85$ in the full sample, with all subsamples having assent frequency $\geq 0.8$. It is interesting that the lowest support for this occurs within the undergraduate and post-2000 samples, with outright rejection passing $0.1$ for post-1980 and post-2000 subsamples. Notably, there is no rejection in the pre-1980 sample, giving an indication that responses to this proposition are evolving. In terms of dissent, Phillips~\cite{phillips2013introduction} and Moore~\cite{moore1998} limit themselves to saying that $\psi$ can determine all the probabilities associated with a system but do not go on to say this is all the information allowed. Whereas, \cite{weinberg2013} openly suggests that interpretive difficulties might be overcome by finding a more general theory to supersede quantum mechanics. 

We also see strong support (a minimum frequency of $0.94$) for the notion that a quantum state describes a single instance (as opposed to an ensemble). Notably, this achieves $0.96$  assent in the full sample and the undergraduate works are near unanimous on this count. Only Peres~\cite{peres1993} dissents from the notion that quantum states describe individual instances. 

In state collapse we find an interesting case. Here, both the assent and rejection frequencies increase as we move from pre-1980 to more modern subsamples. This seems to indicate the point is becoming more polarising over time. The collapse proposition still has subsample assent frequencies above $0.75$ while achieving $0.79$ in the full sample. The rate of rejection for collapse increases substantially when the post-1980 (and post-2000) subsample is compared to the pre-1980 case. However, the assent frequency for collapse also increases from pre to post-1980, perhaps a marker of increased controversy. It is also noteworthy that, in a sample of $\nbooks$ works, there are only three dissenters to the notion that quantum states collapse: Weinberg~\cite{weinberg2013}, Merzbacher~\cite{merzbacher1998}, as well as Binney \& Skinner~\cite{binney2013physics}. Additionally, the topic is not mentioned by Peres, Mandl, M\"uller-Kirsten, and Schiff~\cite{peres1993,mandl2013quantum,muller-kisten2006,schiff1955}. Interestingly, we note that Home~\cite{Home1997} argues that collapse is not essential to characterise the orthodoxy. However, this is strongly disputed by the data gathered here. 

From these results we can also see that an orthodox view requires the state to contain the entire ontological situation and apply to individual particles. This, in turn, requires us to believe that superposition states can be occupied by individual systems. Of the three propositions, only affirmation of collapse reduces slightly if we consider sources after 1980. There are tentative signs of evolution in the consensus here.

Now we turn to propositions on the nature of particles. A strong degree of consensus is clear around the notion that individual particles display wave-like behaviour with assent frequency $0.88$ in the full sample, while maintaining subsample frequencies $\geq 0.78$. The only dissenters here being Weinberg~\cite{weinberg2013}, Binney \& Skinner~\cite{binney2013physics}, and Sakurai~\cite{Sakurai:1167961}. The topic is also not discussed in von Neumann~\cite{vonneumann1955} (perhaps unsurprising given the mathematical focus) and Griffiths~\cite{Griffiths2004Introduction}. Interestingly, the lowered assent in the post-1980 subsample seems to stem from more recent graduate texts rejecting the proposition. There is also notable evolution, in the sense that there are no rejections pre-1980, with these increasing in frequency as we move to post-1980 and post-2000 subsamples. 

Reasonable support also exists for the notion of wave-particle duality with assent $0.78$ in the full sample and subsample frequency $\geq 0.71$. Here only von Neumann~\cite{vonneumann1955}, Binney \& Skinner~\cite{binney2013physics}, and Tong~\cite{tong2025} outright deny this proposition. It is also not mentioned by Weinberg~\cite{weinberg2013}, Mandl~\cite{mandl2013quantum}, Townsend~\cite{townsend2012modern}, Davydov~\cite{davydov1965} and Sakurai~\cite{Sakurai:1167961}. There is discernable evolution based on publication year, rejection is much more common in the post-2000 subsample than either post-1980 or pre-1980 cases.

For the proposition that individual particles being subject to the uncertainty principle, that is, one cannot know a given particle's position and momentum (for example) simultaneously to arbitrary precision, we see more controversy. This is so because the full sample finds assent of $0.85$ and the pre-1980 subsample has nearly $1.0$. However, the post-1980 subsample only registers $0.77$, this then resurges to over $0.9$ in the post-2000 subset. This proposition receives the greatest rejection frequency in the post-1980 subsample. However, it is not rejected at all in the pre-1980 or post-2000 cases.
The works between 1980 and 2000 that cause this are Griffiths~\cite{Griffiths2004Introduction}, Peres~\cite{peres1993}, McIntyre~\cite{mcintyre2012quantum}, Moore~\cite{moore1998}, and Sakurai~\cite{Sakurai:1167961}. This shows some evidence of a change in the consensus over time, but it appears to have been a fluctuation, returning towards the pre-1980 consensus after two decades. Home~\cite{Home1997} does not include this proposition in their conception of the orthodoxy, likely as it is a flawed proposition. As argued in \cite{Popper1982-POPQTA}, the uncertainty principle could not function in the suggested manner. If the measurements from individual particles had the uncertainty principle's scatter, the final ensemble distribution would have a larger uncertainty. Thus, preventing the verification of a lower bound. Despite scepticism on this point, it is clearly wrong to exclude this from the orthodoxy, as the idea is uniformly affirmed by older works and is resurgent in recent texts. It is possible that this could be connected to the rise of interpretations of QM that use ``information'' as an explanatory tool. A common formulation of the uncertainty principle in terms of ``allowed information'' then becomes very tempting.  

It is notable that affirmation of complementarity is generally weak, even the full sample only finds an agreement frequency of $0.52$. This is not due to denial (no one actually rejects it explicitly) but simply never being mentioned in many later texts. This accounts for the sustained drop in assent frequency from pre-1980 to post-1980 and post-2000 samples. Regardless, it cannot have a place in any robust orthodoxy. We also see that the issue of whether particles possess trajectories has become contentious, with frequency $0.51$ in the full sample. This becomes even more contentious in the post-1980 and Wikipedia samples with $0.44$ and $0.39$ respectively. It initially enjoyed an assent frequency $\sim 0.6$ pre-1980 but has dropped to around $\sim 0.4$ in more recent works. The idea is never rejected pre-1980, but rejection appears afterwards and is largely maintained in more recent works. Asserting there are no trajectories still finds agreement in modern works like Kok~\cite{kok2023}, Miller~\cite{miller2010}, Jain~\cite{jain2017}, Basdevant~\cite{basdevant2023}, Konishi \& Paffuti~\cite{Konishi:2009qva}, Cohen-Tannoudji~\cite{Cohen-Tannoudji:101367}, Shankar~\cite{Shankar:102017}, Greiner~\cite{greiner1989}, and Phillips~\cite{phillips2013introduction}. There is also partial agreement in Moore~\cite{moore1998} and Longair~\cite{longair_2013}. However, the mode response is $0$ in the undergraduate, `Wikipedia notable', and post-1980 samples. So there is some indication that this proposition is increasingly marginalised by the orthodoxy.
It might also follow that those who affirm the ``single particle uncertainty'' proposition should, for consistency, agree that particles have no trajectory (at least for resolutions finer than that allowed by uncertainty). However, we have tried to avoid forcing logical consistency on authors where there is some ambiguity or nuance.

Thus, we see that the orthodoxy holds that particles are individually wave-like, have a dual wave-particle nature reflected in their statistics, and that the uncertainty principle applies to measurements performed on single instances of systems (rather than acting only as a statistical deviation over an ensemble).

Next, we turn to propositions around the question of why quantum mechanics predicts only probabilistic results. There is clear consensus, regardless of whether we confine our attention to modern works, that measurement disturbs quantum systems with assent $0.91$ in the full sample, with a minimum of $0.81$ in the graduate case. Notably, the pre-1980 and post-2000 samples have very similar assent frequencies. Indeed, only Peres~\cite{peres1993} actively dissents. Their dissent is justified by the possibility of non-interactive measurement in QM~\cite{Elitzur_1993}. 

There is also support for the idea that quantum systems have no pre-measurement properties, with assent $\sim 0.88$ in the full sample. For time evolution, this has $> 0.9$ support pre-1980 but retains $> 0.8$ in later samples. This proposition is most rejected in the post-1980 sample, at $\approx 0.15$, dropping to $\approx 0.12$ in the post-2000 sample. This is still twice as common as the pre-1980 works. It is rejected by Binney \& Skinner~\cite{binney2013physics}, Mandl~\cite{mandl2013quantum}, and Cohen-Tannoudji~\cite{Cohen-Tannoudji:101367}. Of these, Mandl and Cohen-Tannoudji describe the Born rule as $\vert \psi(x) \vert^2$ giving the probability for a particle being at $x$, instead of merely ``found'' there. 

The notion that probabilities in QM result from the uncertainty principle has always been a minority view it appears, as most authors do not mention anything like it, achieving a maximum assent of $0.4$ in the pre-1980 sample. However, it is not rejected until after 1980 with the rejection rate dropping substantially in the post-2000 works. This coincides with an increase in assent post-2000, where Tong~\cite{tong2025}, Tamvakis~\cite{tamvakis2019}, Longair~\cite{longair_2013}, Konishi \& Paffuti~\cite{Konishi:2009qva}, and M\"uller-Kirsten~\cite{muller-kisten2006} all support the position. In the same sample, only Binney \& Skinner~\cite{binney2013physics} reject it, so the mode response is still to leave it out. 

There is still some support for the notion that the uncertainty principle is a consequence of measurement disturbance, with assent frequency $0.69$. It falls from $0.8$ pre-1980 to around $0.6$ post-1980 and 2000. Thus, providing clear evidence for consensus evolution on this point. Interestingly, the sample with most affirmation for this last proposition is the ``Wikipedia notable'' one, with $0.79$ assent. The only outright rejections here are Rae~\cite{rae2002} and Basdevant~\cite{basdevant2023}, who both note the uncertainties can happily be measured in separate runs of an experiment, so ``disturbance'' need not occur. 

Thus, for the orthodoxy, we do not reveal the values of studied quantum systems but force them to assume definite values (or alter them) via some uncontrolled disturbance attached to the abstract notion of measurement. The ``uncontrolled'' aspect of this is vital to square the probabilistic nature of QM with the proposition that quantum states apply to individual instances of systems. 

In terms of whether the textbooks only discuss the orthodoxy, or trivialise other interpretations, we find this has a $> 0.9$ frequency pre-1980. This drops to $\sim 0.7$ for post-1980 but rises to $\sim 0.85$ for post-2000. Few rejections occur before 1980, with Weinberg~\cite{weinberg2013}, Miller~\cite{miller2010}, Omnes~\cite{Omnes1999-OMNUQM}, Greiner~\cite{greiner1989}, Rae~\cite{rae2002}, and Fong~\cite{fong2013elementary} all providing more nuanced discussions of interpretation. Notably, the rejection rate drops significantly in the post-2000 sample (to below that of even the pre-1980 case). This odd result suggests more recent books talk about other interpretations less, even as the orthodoxy fragments (as we will see below). Perhaps this is typified by Griffiths~\cite{Griffiths2004Introduction}, who views Bell inequality violations as empirically deciding in favour of the orthodoxy.

Next, we can ask whether there are correlations between proposition responses. To do this, we determine the frequency with which answers for different pairs of propositions match. We present the results in Table~\ref{tab:cor1} for the full and post-1980 samples of textbooks. We choose only to display those for which the assent frequency exceeds $0.6$ to determine if there is a strong correlation between orthodox positions. Importantly, the strongest correlation occurs for the pillars of the orthodoxy: ``measurement disturbs systems'', ``no pre-measurement values'', and ``non-ensemble $\psi$'', being correlated $\gtrsim 80$\% of the time. Over all, it is evident that relatively strong pair-wise correlations exist between almost all the propositions with assent frequency $\gtrsim 0.6$. They all correlate strongly to at least one of the pillar propositions as well. This demonstrates that selecting an orthodoxy on this basis yields a consistently affirmed set of propositions. The only exception here is ``uncertainty linked to disturbance'', which generally features in all the weaker correlations. This isn't too surprising, as it falls below a $0.6$ affirmation in post-1980 works. 

\begin{table}[ht!]
	\centering
	\begin{tabular}{|l|l|l|l|}
		\hline
		Proposition 1 & Proposition 2 & Full sample & Post-1980\\
		\hline
		Collapse & Non-ensemble $\psi$ & 0.81(0.79) & 0.81(0.81)\\
		Collapse & Ontologically complete & 0.76(0.74) & 0.74(0.74)\\
		Collapse & Wave-particle duality & 0.69(0.67) & 0.74(0.74)\\
		Collapse & Single particle uncertainty & 0.71(0.69) & 0.67(0.67)\\
		Collapse & Single particle wave-like & 0.79(0.76) & 0.81(0.81)\\
		Collapse & Measurement disturbs system & 0.81(0.76) & 0.81(0.78)\\
		Collapse & No pre-measurement values & 0.81(0.76) & 0.85(0.81)\\
		Collapse & Orthodoxy only & 0.71(0.62) & 0.70(0.59)\\
		Non-ensemble $\psi$ & Ontologically complete & 0.83(0.81) & 0.81(0.78)\\
		Non-ensemble $\psi$ & Wave-particle duality & 0.74(0.74) & 0.70(0.70)\\
		Non-ensemble $\psi$ & Single particle uncertainty & 0.90(0.88) & 0.85(0.81)\\
		Non-ensemble $\psi$ & Single particle wave-like & 0.86(0.83) & 0.81(0.78)\\
		Non-ensemble $\psi$ & Measurement disturbs system & 0.95(0.90) & 0.96(0.89)\\
		Non-ensemble $\psi$ & No pre-measurement values & 0.90(0.88) & 0.89(0.85)\\
		Non-ensemble $\psi$ & Orthodoxy only & 0.83(0.74) & 0.78(0.63)\\
		Ontologically complete & Wave-particle duality & 0.64(0.64) & 0.63(0.63)\\
		Ontologically complete & Single particle uncertainty & 0.79(0.76) & 0.74(0.70)\\
		Ontologically complete & Single particle wave-like & 0.79(0.79) & 0.78(0.78)\\
		Ontologically complete & Measurement disturbs system & 0.81(0.79) & 0.81(0.78)\\
		Ontologically complete & No pre-measurement values & 0.79(0.76) & 0.78(0.74)\\
		Ontologically complete & Orthodoxy only & 0.76(0.69) & 0.74(0.63)\\
		Wave-particle duality & Single particle uncertainty & 0.67(0.67) & 0.59(0.59)\\
		Wave-particle duality & Single particle wave-like & 0.76(0.76) & 0.74(0.74)\\
		Wave-particle duality & Measurement disturbs system & 0.74(0.71) & 0.74(0.70)\\
		Wave-particle duality & No pre-measurement values & 0.76(0.74) & 0.78(0.74)\\
		Wave-particle duality & Orthodoxy only & 0.64(0.57) & 0.59(0.48)\\
		Single particle uncertainty & Single particle wave-like & 0.83(0.81) & 0.78(0.74)\\
		Single particle uncertainty & Measurement disturbs system & 0.86(0.83) & 0.81(0.78)\\
		Single particle uncertainty & No pre-measurement values & 0.81(0.76) & 0.74(0.67)\\
		Single particle uncertainty & Orthodoxy only & 0.79(0.74) & 0.70(0.63)\\
		Single particle wave-like & Measurement disturbs system & 0.83(0.81) & 0.81(0.78)\\
		Single particle wave-like & No pre-measurement values & 0.86(0.83) & 0.85(0.81)\\
		Single particle wave-like & Orthodoxy only & 0.79(0.74) & 0.74(0.67)\\
		Measurement disturbs system & No pre-measurement values & 0.86(0.81) & 0.85(0.78)\\
		Measurement disturbs system & Orthodoxy only & 0.81(0.74) & 0.78(0.67)\\
		No pre-measurement values & Orthodoxy only & 0.79(0.69) & 0.74(0.59)\\
		\hline
	\end{tabular}
	\caption{The frequency with which answers for two propositions match in the full and post-1980 samples. Values in brackets indicate those without inferred responses.}
	\label{tab:cor1}
\end{table}

\begin{figure}[htbp]
	\centering
	\resizebox{0.9\hsize}{!}{\includegraphics{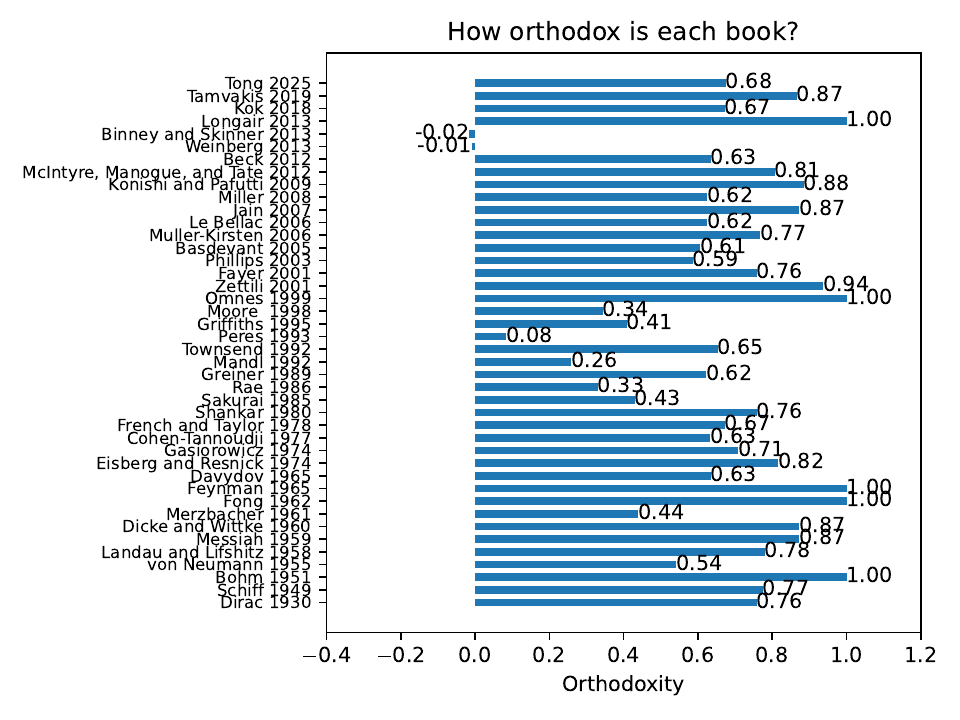}}
	\caption{The ``orthodoxity'' or degree to which each studied source agrees with the averaged responses to each proposition. This is ordered by date of first publication.}
	\label{fig:orthodox}
\end{figure}

Finally, we have Figure~\ref{fig:orthodox}. This displays the ``orthodoxity'' of each book, i.e. how close they are to the averaged responses over the full sample. This is determined following
\begin{equation}
	O = \frac{\sum_i^N w_i a_i}{\sum_i^N w_i} \; ,
\end{equation}
where the sum runs over the propositions, $w_i$ is the average response value, and $a_i$ is that of the reference work in question. Note that we exclude propositions that do not have a full-sample assent frequency $>0.6$ as well as ``orthodoxy only''. As illustrated above, this selection criterion results in set of well-correlated propositions.

The average orthodoxity is $66$\% with standard deviation $26$\% indicating there isn't a very robust degree of orthodoxity on average. This includes inferred values, but, this effect is small. Note that these averages are robust to increasing the frequency inclusion threshold to $0.7$ as well. For convenience, we order Fig.~\ref{fig:orthodox} by year of first publication. This allows us to determine if there are any trends in the orthodoxy. Early works adhere very closely to the full-sample orthodoxy, with those before 1980 having an average orthodoxity of $77$\% with standard deviation $16$\%. The period around the early 90's shows a fluctuation of far lower agreement with the orthodoxy. This results in post 1980 works having an average orthodoxity of $60$\% and standard deviation $28$\%. The substantial nature of this change from the earlier consensus is most vividly illustrated by the post-1980 standard deviation almost doubling that of pre-1980, despite a $73$\% increase in sample size. The 90's dip in the orthodoxity is constituted by the works: Peres~\cite{peres1993}, Moore~\cite{moore1998}, Griffiths~\cite{Griffiths2004Introduction}, Mandl~\cite{mandl2013quantum}, Rae~\cite{rae2002}, and Sakurai~\cite{Sakurai:1167961}. Of these, Mandl is here mainly due to omission, rather than contesting orthodox positions. If we replace Mandl with Weinberg~\cite{weinberg2013}, we form a set of all the authors who deny more than one proposition. However, there is no consistent pattern in these denials. Interestingly, the average orthodoxity rises to $66 \pm 28$\% as we transition to the post-2000 sample. Combined with the large drop-off in denials of ``orthodoxy only'', this suggests there is some resurgent support for the orthodoxy. However, the standard deviation remaining nearly double the pre-1980 value indicates that fragmentation continues. 

Overall, there appears to be a post-1980 divergence from the earlier orthodoxy. This being said, it resembles a fragmentation, rather than a new consensus emerging, due to the lack of consistent patterns in the denial of previously orthodox positions. However, it must be admitted that many modern works adhere quite closely to the interpretation put forward by their predecessors. This suggests the possibility that our present epoch is the start of a collapse in confidence in the orthodoxy. A survey of working physicists has recently demonstrated that the community is fragmented on the interpretation of quantum mechanics~\cite{survey2025}, notably more so than a previous survey in 2016~\cite{sivasundaram2016surveyingattitudesphysicistsconcerning}. This could well signal an increase in divergence from the previous quantum orthodoxy in future textbooks. Somewhat complicating such an idea is the tendency of more orthodox post-1980 works to cite Bell's theorem as a vindication of many aspects of the orthodoxy, perhaps accounting for the apparent resurgence in orthodoxity after the late 1990's. The timeline notably aligns with the emergence of ``Copenhagen+'' interpretations like quantum Bayesianism (see \cite{timpson_quantum_2008}).

For additional works that were not included in the sample we find the following orthodoxity measures: Ballentine~\cite{ballentine1998quantum} $-0.88$, Home~\cite{Home1997} 0.3, Hannabuss~\cite{Hannabuss:1999fa} 0.16, and Nielsen \& Chuang~\cite{nielsen2010} 0.3. The last of these, Nielsen \& Chuang~\cite{nielsen2010}, only shows a lower orthodoxity by virtue of never mentioning the topic of waves. Their strong endorsement of Peres~\cite{peres1993} and Sakurai~\cite{Sakurai:1167961} might indicate little sympathy with this aspect of the orthodoxy, but the clarity of this is marred by also endorsing Cohen-Tannoudji~\cite{Cohen-Tannoudji:101367}. Interestingly, Home's characterisation of the orthodoxy~\cite{Home1997} appears incomplete at $0.3$. Additionally, this is only when one concentrates on the aspects they attributed to the non-ensemble case. Home considers ensemble views as part of the orthodoxy on the basis that they view the formalism as being complete (there are no supplemental entities needed to explain what is going on). This principle as the central pillar of the orthodoxy is apriori dubious. This is because the notion of collapse is a supplement to the basic mathematical formalism~\cite{Wallace2019,ballentine1998quantum,weinberg2013}, in the form of an additional physical/quasi-physical process with attendant projection postulate. Since Copenhagen-like interpretations hinge so heavily on collapse, their popularity implies that Home's primary condition of orthodoxy becomes more doubtful. Interestingly, Ballentine~\cite{ballentine1998quantum}, who would express an orthodox view according to Home, achieves an orthodoxity of $-0.88$ here, dissenting from (or not addressing) every proposition considered. This is because Home's rationale for including ensemble views in the orthodoxy may not be well justified. Home argues that, to enjoy a complete formalism, the ensemble must be supplemented by hidden variables that explain the assignment of definite values to individual realisations. However, an ensemble view only requires statistical completeness, $\psi$ can answer all statistical questions, not true ontological completeness. In this regard Home forces ensemble views to be able to determine the behaviour of single instances, in contradiction with their basic premise. In addition, no such definite assignment is required of other statistical theories in science (which usually exist as a limit of some other, more detailed theory). We do not require statistical mechanics to explain how velocities are assigned to the molecules in a gas, for instance.

\begin{figure}[htbp]
	\centering
	\resizebox{0.9\hsize}{!}{\includegraphics{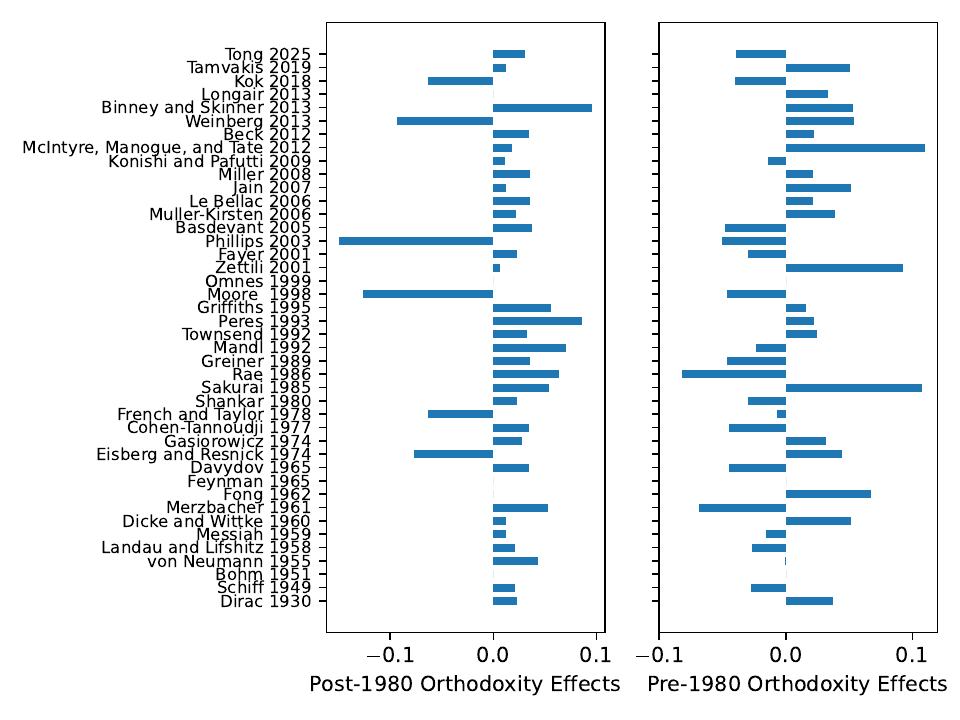}}
	\caption{The changes in ``orthodoxity'' that result if we use different samples to define the orthodoxy.}
	\label{fig:orthodox_changes}
\end{figure}

We can study how time period effects the orthodoxy by changing the orthodoxity weights to those of a subsample. We display the changes in orthodoxy for both pre and post-1980 selections in Fig.~\ref{fig:orthodox_changes}. Importantly, the pre-1980 case includes three extra propositions: ``no trajectories'', ``uncertainty linked to disturbance'', and ``complementarity'', as all pass the $0.6$ assent threshold in that subsample only. For both our modified orthodoxies, the mean changes from the full sample are small, being $0.012$ and $0.006$ respectively. This indicates a small preference for the post-1980 orthodoxy, with several authors becoming around $10$\% more orthodox in this case but very few reducing. In contrast, the pre-1980 case has many decrements, despite larger gains. Interestingly, the largest post-1980 decreases are all for post-1980 works. Whereas, the largest gains for a pre-1980 orthodoxy are from post-1980 works as well. This indicates there are definite hints of evolution in the orthodoxy. However, there are no significant correlations between the changes (for any of our temporal subsamples), indicating that a new consistent orthodoxy is yet to emerge.

We can now ask how far the revealed orthodoxy diverges from the idea that it can be defined entirely by the ontological primacy of the quantum state (as advocated by Home~\cite{Home1997}). The first element that brings this into question is the ``non-ensemble $\psi$'' proposition. This is a vital part of the orthodoxy (stronger even than ontological completeness) and it does not follow from $\psi$ being ontologically complete. There being no reason to preclude a fundamentally indeterminist viewpoint. If this is the case, then there are a cascade of propositions that must follow, as they all depend on ``non-ensemble $\psi$'' for their relevance. These are ``single particles are wave-like'', ``states collapse on measurement'', ``measurement inherently disturbs the system'', and ``measurement creates observed values''.  This indicates that a single proposition reduction of the orthodoxy is unviable. The two main pillars that are empirically revealed are ``measurement disturbs systems'' and ``states do not just describe ensembles''. Notably, the significance of consensus around ontological completeness has declined for modern works, as a result of some authors~\cite{phillips2013introduction,moore1998} instead positing ``statistical completeness'' or openly suggesting a more general theory should be found~\cite{weinberg2013}.

\section{Conclusion}\label{sec13}

When we consider a sample of $\nbooks$ textbooks, aimed at both graduate and undergraduate students of quantum mechanics, we find evidence of robust orthodoxy that has begun to fragment. This is powerfully illustrated by post-1980 works having an average orthodoxity of $60$\%, down from a pre-1980 value of $77$\%. The standard deviation almost doubles from $16$\% to $28$\% when moving from pre to post-1980 samples. For the full sample, the average orthodoxity of the sources is $66$\% with a standard deviation of $25$\%. Over time, there has been less support for the notions that individual particles are wave-like, and that the quantum state is ontologically complete. There are also more significant reductions in support for the notions that the uncertainty principle applies to individual particles, quantum uncertainty results from measurement disturbance, particles do not have trajectories, and complementarity. The reduction in orthodoxy over time mirrors recent surveys of physicists attitudes in quantum mechanics~\cite{survey2025}. Perhaps indicating the start of opinion fragmenting away from the orthodoxy, as there are no obvious patterns for groups of propositions being denied together. 

For the full sample of works, the revealed orthodoxy strongly resembles many aspects of Copenhagen-like frameworks. Particularly, there is powerful agreement (assent frequency $0.91$) that measurement inherently disturbs quantum systems, quantum systems have no pre-measurement values ($0.91$ assent), single instances of systems are individually described by a quantum state, rather than as an ensemble (assent frequency $0.96$), state collapse happens (assent frequency $0.79$), that individual particles are wave-like (assent frequency $0.88$), and that quantum states are ontologically complete (assent frequency $0.85$). These positions are further supplemented by the existence of a wave-particle duality (assent frequency $0.76$) and that individual particle measurements are subject to the uncertainty principle (assent frequency $0.85$). More contentious, but still popular, is the notion that the uncertainty principle arises from measurement disturbance (assent frequency $0.67$). 

It is clear that the orthodoxy favours the view of the quantum state as a complete description of the actual state of affairs, rather than the more instrumental view attributed to Copenhagen by authors like Stapp~\cite{stapp1972}. In this regard, the views of the orthodoxy espoused by Home~\cite{Home1997} and Jammer~\cite{Jammer1974-JAMTPO-10} are vindicated. The work most embodying Stapp's version would be Peres~\cite{peres1993}, with similarities in Mandl~\cite{mandl2013quantum}, Nielsen \& Chuang~\cite{nielsen2010}, Hannabuss~\cite{Hannabuss:1999fa}, and Sakurai~\cite{Sakurai:1167961}. However, it is not clear that all the logical consequences of Home's view (that the orthodoxy is entirely characterised by ontological completeness) are believed by the physics community (and it is clear there are wave-related addenda). Thus, the orthodoxy's characterisation requires more nuance than being distilled to a single point about the status of quantum states. It is clear that, examining all the propositions in this work, there is no room in the orthodoxy for ensemble views, entirely in opposition to the stance taken by Home. In fact, it is a strong requirement of the orthodoxy that quantum states describe single instances, rather than ensembles. 

It is also interesting to note that respondents in \cite{sivasundaram2016surveyingattitudesphysicistsconcerning} characterise the Copenhagen interpretation's most salient features as ``collapse'' (77\%), ``complementarity'' (71\%), ``indeterminism'' (46\%), and ``correspondence'' (43\%). This is quite different to the orthodoxy deduced from our sample of textbooks. Where complementarity is far less important, indeterminism is more strongly supported (non-ensemble $\psi$ + no pre-measurement values), and collapse is popular but not absolutely primary. This suggests that, although physicists might predominantly report subscribing to the Copenhagen interpretation, the prevailing community beliefs do not fully conform to their own characterisation of this viewpoint. 

To an orthodox quantum mechanic, individual particles exist in wave-like superpositions, when left to their own devices, and assume their observed properties (and appear as particles) when forced to by outside interference. When this happens, the individual particle's properties are subject to the uncertainty principle as a form of measurement uncertainty and the state is inevitably changed by the uncontrolled disturbance inherent in measurement. 

\backmatter

\bmhead{Acknowledgements}
GB acknowledges funding from the National Research Foundation of South Africa under the incentive funding for rated researchers program. 

\bmhead{Declarations}
Data used in this work is available on reasonable request. The authors declare no conflicts of interest.

\bibliography{qm_lit_survey}

@book{Omnes1999-OMNUQM,
	year = {1999},
	title = {Understanding Quantum Mechanics},
	publisher = {Princeton University Press},
	author = {Roland Omn\`{e}s},
	address={New Jersey, USA}
}

@book{davydov1965,
	authyor = "A. S. Davydov",
	year = 1976,
	edition="2nd",
	title="Quantum mechanics",
	publisher="Pergamon Press",
	address="Oxford,UK",
}

@book{jain2017,
	year = 2017,
	author = "Mahesh C. Jain",
	title = "Quantum mechanics: a textbook for undergraduates",
	publisher = "PHI Learning",
	address = "Delhi, India",
	eition = "2nd",
}

@book{miller2010,
	year = 2010,
	author="David A. B Miller",
	title="Quantum mechanics for scientists and engineers",
	publisher = "Cambridge University Press",
	address="New York, USA",
}

@book{basdevant2023,
	year = 2023,
	publisher="Springer Nature",
	title = "Lectures on quantum mechanics",
	address="Switzerland",
	edition="3rd",
	author = "Jean-Louis Basdevant",
}

@book{beck2012,
	author = "Mark Beck",
	publisher="Oxford University Press",
	address="Oxford,UK",
	title="Quantum mechanics: theory and experiment",
	year = 2012,
}

@book{fayer2001,
	year = 2001,
	author = "M. D. Fayer",
	publisher = "Oxford University Press",
	address = "Oxford, UK",
	title = "Elements of quantum mechanics",
}

@book{kok2023,
	year = "2023",
	author = "Pieter Kok",
	title = "A first introduction to quantum physics",
	publisher = "Springer Nature",
	address="Switzerland",
	edition="2nd",
}

@book{merzbacher1998,
	year = 1998,
	author = "Eugen Merzbacher",
	title = "Quantum mechanics",
	publisher="John Wiley \& Sons",
	address = "New York, USA",
	edition="3rd",
}

@book{binney2013physics,
	title={The Physics of Quantum Mechanics},
	author={Binney, J. and Skinner, D.},
	isbn={9780199688579},
	lccn={2013940021},
	year={2013},
	publisher={Oxford University Press},
	address = "Oxford, UK",
}

@book{zettili2001,
	author = "Zettili, N.",
	year = 2001,
	publisher = "Wiley",
	title = "Quantum mechanics: concepts and applications",
	address = "Chichester",
	}

@book{vonneumann1955,
	author = "von Neumann, J.",
	year =1955,
	title = "Mathematical foundations of quantum mechanics",
	publisher = "Princeton University Press",
	address = "Princeton, New Jersey",}

@book{dicke1960,
	author = "Dicke, R. H. and Wittke, J. P.",
	year = 1960,
	title = "Introduction to quantum mechanics",
	publisher = "Addison-Wesley",
	address="Reading, Massachusetts",}

@book{schiff1955,
	author = "Schiff, L. I.",
	year = 1955,
	title = "Quantum mechanics",
	publisher = "McGraw-Hill Book company",
	address= "New York, USA",}

@book{muller-kisten2006,
	title = "Introduction to quantum mechanics: Schrodinger equation and path integral",
	year = 2006,
	author = {M\"uller-Kirsten, H. J.},
	publisher = "World Scientific Publishing",
	address = "Singapore",
	 }

@book{longair_2013, place={Cambridge}, title={Quantum Concepts in Physics: An Alternative Approach to the Understanding of Quantum Mechanics}, DOI={10.1017/CBO9781139062060}, publisher={Cambridge University Press}, author={Longair, Malcolm}, year={2013}, address={Cambridge, UK},}

@book{Jammer1974-JAMTPO-10,
	address = {New York,},
	author = {Max Jammer},
	editor = {Max Jammer},
	publisher = {Wiley},
	title = {The Philosophy of Quantum Mechanics},
	year = {1974}
}

@book{french1978,
	author = "French, A. P. and Taylor, E. F.",
	title = "An introduction to quantum physics",
	publisher = "Thomas Nelson and Sons Ltd",
	year = "1979",
	address = "Middlesex, England",
}

@book{weinberg2013,
	author = "Weinberg, S.",
	title = "Lectures on quantum mechanics",
	publisher = "Cambridge university press",
	year = "2013",
	address = "New York, USA",}

@book{eisberg1974,
	author = "Eisberg, R. and Resnick, R.",
	year = "1974",
	publisher = "John Wiley \& Sons",
	address = "New York, USA",
	title = "Quantum physics of atoms, molecules, solids, nuclei and particles",
}

@article{Elitzur_1993,
	title={Quantum mechanical interaction-free measurements},
	volume={23},
	ISSN={1572-9516},
	url={http://dx.doi.org/10.1007/BF00736012},
	DOI={10.1007/bf00736012},
	number={7},
	journal={Foundations of Physics},
	publisher={Springer Science and Business Media LLC},
	author={Elitzur, Avshalom C. and Vaidman, Lev},
	year={1993},
	month=jul, pages={987–997} }

@article{bohm1982,
author = {D Bohm and B J Hiley},
journal = {Foundations of physics},
year = 1982,
volume=12,
pages=1001,
title ={The de Broglie pilot wave theory and the further development of new insights arising out of it}}

@article{Rovelli_1996,
	title={Relational quantum mechanics},
	volume={35},
	ISSN={1572-9575},
	url={http://dx.doi.org/10.1007/BF02302261},
	DOI={10.1007/bf02302261},
	number={8},
	journal={International Journal of Theoretical Physics},
	publisher={Springer Science and Business Media LLC},
	author={Rovelli, Carlo},
	year={1996},
	month=aug, pages={1637–1678} }

@InCollection{sep-qm-consistent-histories,
	author       =	{Griffiths, Robert},
	title        =	{{The Consistent Histories Approach to Quantum Mechanics}},
	booktitle    =	{The Stanford Encyclopedia of Philosophy},
	address = {Stanford, USA},
	editor       =	{Edward N. Zalta},
	howpublished =	{\url{https://plato.stanford.edu/archives/sum2019/entries/qm-consistent-histories/}},
	year         =	{2019},
	edition      =	{{S}ummer 2019},
	publisher    =	{Metaphysics Research Lab, Stanford University}
}

@article{Fuchs_2013,
	title={Quantum-Bayesian coherence},
	volume={85},
	ISSN={1539-0756},
	url={http://dx.doi.org/10.1103/RevModPhys.85.1693},
	DOI={10.1103/revmodphys.85.1693},
	number={4},
	journal={Reviews of Modern Physics},
	publisher={American Physical Society (APS)},
	author={Fuchs, Christopher A. and Schack, Rüdiger},
	year={2013},
	month=dec, pages={1693–1715} }

@article{hanson1959,
	author = {Hanson, Norwood Russell},
	title = "{Copenhagen Interpretation of Quantum Theory}",
	journal = {American Journal of Physics},
	volume = {27},
	number = {1},
	pages = {1-15},
	year = {1959},
	month = {01},
	issn = {0002-9505},
	doi = {10.1119/1.1934739},
	url = {https://doi.org/10.1119/1.1934739},
	eprint = {https://pubs.aip.org/aapt/ajp/article-pdf/27/1/1/11942472/1\_1\_online.pdf},
}

@article{stapp1972,
	author = {Stapp, Henry Pierce},
	title = "{The Copenhagen Interpretation}",
	journal = {American Journal of Physics},
	volume = {40},
	number = {8},
	pages = {1098-1116},
	year = {1972},
	month = {08},
	issn = {0002-9505},
	doi = {10.1119/1.1986768},
	url = {https://doi.org/10.1119/1.1986768},
	eprint = {https://pubs.aip.org/aapt/ajp/article-pdf/40/8/1098/10114905/1098\_1\_online.pdf},
}

@incollection{feyeraband1962,
	ISBN = {9780822931003},
	URL = {http://www.jstor.org/stable/jj.5973228.10},
	author = {P. K. Feyerabend},
	editor={R G Colodny},
	address={Pittsburgh, USA},
	booktitle = {Frontiers of Science and Philosophy},
	pages = {189--284},
	publisher = {University of Pittsburgh Press},
	title = {Problems of Microphysics},
	year = {1962}
}

@incollection{shimony1993,
	author = {Abner Shimony},
	title={Physical and philosophical issues in the Bohr-Einstein debate},
	booktitle={Foundations Of Modern Physics 1992 - Proceedings Of The Symposium},
	editor={Laurikainen, K.V. and Montonen, C.},
	isbn={9789814553315},
	address={Singapore},
	year={1993},
	pages={79--96},
	publisher={World Scientific Publishing Company}
}

@incollection{Wallace2019,
	author="Wallace, David",
	editor="Cordero, Alberto",
	title="What is Orthodox Quantum Mechanics?",
	bookTitle="Philosophers Look at Quantum Mechanics",
	year="2019",
	publisher="Springer International Publishing",
	address="Cham",
	pages="285--312",
	isbn="978-3-030-15659-6",
}

@book{nielsen2010,
	year = 2010,
	title = {Quantum computation and quantum information},
	author = {M. A. Nielsen and I. L. Chuang},
	publisher = {Cambridge University Press},
	address = {Cambridge, UK},
}

@book{moore1998,
	year = 2017,
	title = {Six ideas that shaped physics: particles behave like waves},
	author = {T. A. Moore},
	publisher = {McGraw-Hill Education},
	address = {New York, USA},
}

@book{bohm1951,
	year = {1989},
	title = {Quantum Theory},
	publisher = {Dover publications},
	address={Mineola, USA},
	author = {David Bohm}
}

@book{Sakurai:1167961,
	author        = "Sakurai, Jun John",
	title         = "Modern quantum mechanics",
	publisher     = "Addison-Wesley",
	address       = "Reading, MA",
	year          = "1985",
}

@book{tong2025,
	author = "David Tong",
	year = 2025,
	title = "Quantum Mechanics: Lectures on Theoretical Physics",
	publisher = "Cambridge University Press",
	address="Cambridge, UK",
}

@article{survey2025,
	author = "Elizabeth Gibney",
	journal = "Nature",
	volume =  643, 
	pages = {1175-1179},
	year =  2025,
}

@article{timpson_quantum_2008,
	title = {Quantum {Bayesianism}: {A} study},
	volume = {39},
	issn = {1355-2198},
	shorttitle = {Quantum {Bayesianism}},
	url = {https://www.sciencedirect.com/science/article/pii/S1355219808000257},
	doi = {10.1016/j.shpsb.2008.03.006},
	number = {3},
	urldate = {2025-07-31},
	journal = {Studies in History and Philosophy of Science Part B: Studies in History and Philosophy of Modern Physics},
	author = {Timpson, Christopher Gordon},
	month = sep,
	year = {2008},
	pages = {579--609},
}

@misc{sivasundaram2016surveyingattitudesphysicistsconcerning,
	title={Surveying the Attitudes of Physicists Concerning Foundational Issues of Quantum Mechanics}, 
	author={Sujeevan Sivasundaram and Kristian Hvidtfelt Nielsen},
	year={2016},
	eprint={1612.00676},
	archivePrefix={arXiv},
	primaryClass={physics.hist-ph},
	url={https://arxiv.org/abs/1612.00676}, 
}

@book{messiah1999quantum,
title={Quantum Mechanics},
author={Messiah, A.},
number={v. 2},
isbn={9780486409245},
lccn={99055362},
series={Dover books on physics},
year={1999},
publisher={Dover Publications},
address={New York, USA}
}

@book{greiner1989,
title={Quantum Mechanics: An Introduction},
author={Greiner, W.},
isbn={9783540674580},
lccn={00046345},
series={Physics and Astronomy},
year={2001},
publisher={Springer},
address={Berlin, Germany}
}

@book{Konishi:2009qva,
author = "Konishi, Kenichi and Paffuti, Giampiero",
title = "Quantum mechanics: A new introduction",
isbn = "978-0-19-956026-4, 978-0-19-956027-1",
publisher = "Oxford Univ. Pr.",
address = "Oxford, UK",
year = "2009"
}

@book{mandl2013quantum,
title={Quantum Mechanics},
author={Mandl, F.},
isbn={9780471931553},
series={Manchester Physics Series},
year={1992},
publisher={Wiley},
address={Chichester, UK}
}

@book{Hannabuss:1999fa,
author = "Hannabuss, K.",
title = "An introduction to quantum theory",
year = "1997",
publisher={Oxford University Press},
address={Oxford, UK}
}

@book{Shankar:102017,
author        = "Shankar, Ramamurti",
title         = "Principles of quantum mechanics",
publisher     = "Springer Science+Business Media",
address       = "New York, USA",
year          = "2008",
}

@book{mcintyre2012quantum,
title={Quantum Mechanics},
author={McIntyre, D. and Manogue, C.A. and Tate, J.},
isbn={9780321850003},
year={2012},
publisher={Pearson Education},
address={London, UK}
}

@book{feynman1977feynman,
	title={The Feynman Lectures on Physics},
	author={Feynman, R.P. and Leighton, R.B. and Sands, M.},
	number={v. 3},
	isbn={9780201021189},
	lccn={63207177},
	year={2010},
	publisher={Basic books},
	address={New York, USA}
}

@book{Dirac1930-DIRTPO,
	address = {Oxford, UK},
	author = {Paul Adrien Maurice Dirac},
	editor = {},
	publisher = {Clarendon Press},
	title = {The Principles of Quantum Mechanics},
	year = {1930}
}

@book{rae2002,
	author= "A. Rae",
	title = "Quantum mechanics",
	edition = "4th",
	publisher = "Institute of Physics",
	address = {Bristol, UK},
	year = {2002},
}

@book{Popper1982-POPQTA,
	address = {New York, USA},
	author = {Karl Raimund Popper},
	editor = {},
	publisher = {Routledge},
	title = {Quantum Theory and the Schism in Physics},
	year = {1982}
}

@book{Griffiths2004Introduction,
author = {David Griffiths},
publisher = {Pearson Prentice Hall},
title = {Introduction to Quantum Mechanics (2nd International Edition)},
year = 2005,
address={New Jersey, USA}
}

@book{tamvakis2019,
author = {Kyriakos Tamvakis},
title = {Basic quantum mechanics},
publisher={Springer Nature},
address={Switzerland},
year=2019}

@book{bellac2011quantum,
author={Le Bellac,Michel},

year={2006},

title={Quantum physics},

publisher={Cambridge University},

address={Cambridge},

isbn={9780521852777;0521852773;},

language={English},
}

@book{townsend2012modern,
title={A Modern Approach to Quantum Mechanics},
author={Townsend, J.},
isbn={9781891389788},
lccn={2011049655},
year={2012},
publisher={University Science Books},
address={California, USA}
}

@book{peres1993,
title = {Quantum theory: concepts and methods},
author={Peres, A},
year={1993},
publisher={Kluwer Academic},
address = {Dordrecht, Germany}
}

@book{fong2013elementary,
title={Elementary Quantum Mechanics},
author={Fong, P.},
year={1962},
publisher={Addison-Wesley},
address={Massachusetts, USA}
}

@book{phillips2013introduction,
title={Introduction to Quantum Mechanics},
author={Phillips, A.C.},
series={Manchester Physics Series},
year={2003},
publisher={Wiley},
address={New Jersey, USA}
}

@book{gasiorowicz2007quantum,
title={Quantum Physics, 3rd Ed},
author={Gasiorowicz, S.},
year={2003},
publisher={Wiley},
address={New Jersey, USA}
}

@book{Cohen-Tannoudji:101367,
author        = {Cohen-Tannoudji, Claude and Diu, Bernard and Lal\"e Franck},
title         = {Quantum mechanics; 1st ed.},
publisher     = "Wiley",
address       = "New York, USA",
year          = "1977",
note          = "Trans. of : Mécanique quantique. Paris : Hermann, 1973",
}

@book{Landau1981Quantum,
author = {Landau, L. D. and Lifshitz, L. M.},
edition = 3,
month = jan,
publisher = {Butterworth-Heinemann},
address={Oxford, UK},
title = {Quantum Mechanics Non-Relativistic Theory, Third Edition: Volume 3},
year = 1977
}

@book{ballentine1998quantum,
title={Quantum Mechanics: A Modern Development},
author={Ballentine, L.E.},
isbn={9789810241056},
lccn={98012768},
year={1998},
publisher={World Scientific},
address={Singapore}
}

@book{Home1997,
publisher = {Springer Verlag},
year = {1997},
title = {Conceptual Foundations of Quantum Physics: An Overview From Modern Perspectives},
author = {Dipankar Home},
address={Berlin, Germany}
}

\end{document}